\documentclass[sigconf, nonacm]{acmart}
\expandafter\def\expandafter\normalsize\expandafter{%
    \normalsize
    \setlength\abovedisplayskip{3pt}
    \setlength\belowdisplayskip{3pt}
    \setlength\abovedisplayshortskip{3pt}
    \setlength\belowdisplayshortskip{3pt}
}

\newcommand\vldbdoi{10.14778/3430915.3430921}
\newcommand\vldbpages{307 - 319}
\newcommand\vldbvolume{14}
\newcommand\vldbissue{3}
\newcommand\vldbyear{2021}
\newcommand\vldbauthors{\authors}
\newcommand\vldbtitle{\shorttitle} 
\newcommand\vldbavailabilityurl{https://github.com/sunlab-osu/TURL}
\newcommand\vldbpagestyle{empty} 

\newif\ifdebugdoc\debugdoctrue

\ifdebugdoc
\newcommand{\fyi}[1]{\footnote{\textcolor{blue}{fyi:#1}}}

\newcommand{\del}[1]{\textcolor{blue}{\st{#1}}}
\newcommand{\outline}[1]{\textbf{\colorbox{yellow}{Outline:}\textcolor{red}{#1.}}}

\newcommand{\jie}[1]{\footnote{\colorbox{yellow}{Jie:} #1.}}

\else
\newcommand{\fyi}[1]{}

\newcommand{\add}[1]{#1}
\newcommand{\del}[1]{}
\newcommand{\jie}[1]{}
\newcommand{\outline}[1]{}
\fi

\newcommand{\nop}[1]{}
\usepackage{balance}
\usepackage{url}
\usepackage{enumitem}
\usepackage{amsmath}
\usepackage{booktabs}
\usepackage{multirow}
\usepackage{graphicx}
\usepackage{tabularx}
\newcolumntype{Y}{>{\centering\arraybackslash}X}
\usepackage{soul}
\usepackage{amsthm}
\usepackage{subcaption}

\usepackage{makecell}
\usepackage{cases}
\usepackage{scalerel}

\newcommand{\model}{\textsf{TURL} }
\newcommand{\modelnospace}{\textsf{TURL}}

\newcommand{\hs}[1]{\textcolor{blue}{Huan: #1}}
\newcommand{\xd}[1]{\textcolor{ForestGreen}{Xiang: #1}}
\newcommand{\add}[1]{\textcolor{brown}{#1}}
\newcommand{\al}[1]{\textcolor{red}{Alyssa: #1}}
\newcommand{\cy}[1]{\textcolor{Bittersweet}{Cong: #1}}
\newcommand{\ww}[1]{\textcolor{BlueGreen}{Will: #1}}
\newcommand{\revision}[2]{#2}

\usepackage[font={small}]{caption}

\begin{document}
%\title{Unsupervised Representation Learning on Web Tables}
\title{TURL: Table Understanding through Representation Learning}

%%
%% The "author" command and its associated commands are used to define the authors and their affiliations.

\author{Xiang Deng}
\authornote{Corresponding authors.}
\affiliation{%
  \institution{The Ohio State University}
  \city{Columbus}
  \state{Ohio}
}
\email{deng.595@buckeyemail.osu.edu}

\author{Huan Sun}
\authornotemark[1]
\affiliation{%
  \institution{The Ohio State University}
  \city{Columbus}
  \state{Ohio}
}
\email{sun.397@osu.edu}

\author{Alyssa Lees}
\affiliation{%
  \institution{Google Research}
  \city{New York}
  \state{NY}
}
\email{alyssalees@google.com}

\author{You Wu}
\affiliation{%
  \institution{Google Research}
  \city{New York}
  \state{NY}
}
\email{wuyou@google.com}

\author{Cong Yu}
\affiliation{%
  \institution{Google Research}
  \city{New York}
  \state{NY}
}
\email{congyu@google.com}

\begin{abstract}
{Relational tables on the Web store a vast amount of knowledge. Owing to the wealth of such tables, there has been tremendous progress on a variety of tasks in the area of table understanding. However, existing work generally relies on heavily-engineered task-specific features and model architectures. In this paper, we present \modelnospace, a novel framework that introduces the pre-training/fine-tuning paradigm to relational Web tables. During pre-training, our framework learns deep contextualized representations on relational tables in an unsupervised manner. Its universal model design with pre-trained representations can be applied to a wide range of tasks with minimal task-specific fine-tuning.

Specifically, we propose a structure-aware Transformer encoder to model the row-column structure of relational tables, and present a new Masked Entity Recovery (MER) objective for pre-training to capture the semantics and knowledge in large-scale unlabeled data.
We systematically evaluate \model with a benchmark consisting of 6 different tasks for table understanding (e.g., relation extraction, cell filling). We show that \model generalizes well to all tasks and substantially outperforms existing methods in almost all instances.\nop{ Our source code, benchmark, as well as pre-trained models will be available online to facilitate future research.\footnote{\url{https://github.com/sunlab-osu/TURL}}}
}

\nop{\hs{how about this title: TURL: Table-based Unsupervised Representation Learning? TURL is the name of our model which can be pre-trained on web tables and fine-tuned on downstream tasks.
~\\ 
Alternatives: ~\\
1. PreTrainTable: Deep Contextualized Representation Learning on Tables (more intuitive; easier to remember)~\\
2. TableBert: Deep Contextualized Representation Learning on Tables ~\\
3. Deep Contextualized Representation Learning on Tables: A Pre-training--Fine-tuning Paradigm and a Table Understanding Benchmark
}

\cy{I like TURL as a short name, but perhaps for Table Understanding through Representation Learning}

Semi-structured tables, especially relational tables, embed a vast amount of information on the Web. In this paper, we present a new table representation model \model to learn general table representation from large-scale table corpora. Unlike previous works that use manually designed features or embeddings learned by shallow Word2Vec to represent tables, \model captures semantic, structure and knowledge information in relational tables simultaneously via pre-training a structure-aware transformer with table-specific objectives. \ww{Knowledge is a very broad term.  The focus of this work is entity relations, not numerical facts.  This should be made clear in intro.} The learned representation can be easily incorporated into existing models for table-driven tasks. Evaluation results on three representative tasks: row population, cell filling and attribute discovery show that table representation generated by XXX brings significant improvement to baselines. \hs{to revise later once intro is done}
}
\end{abstract}

\maketitle

\pagestyle{\vldbpagestyle}
\begingroup\small\noindent\raggedright\textbf{PVLDB Reference Format:}\\
\vldbauthors. \vldbtitle. PVLDB, \vldbvolume(\vldbissue): \vldbpages, \vldbyear.\\
\href{https://doi.org/\vldbdoi}{doi:\vldbdoi}
\endgroup
\begingroup
\renewcommand\thefootnote{}\footnote{\noindent
This work is licensed under the Creative Commons BY-NC-ND 4.0 International License. Visit \url{https://creativecommons.org/licenses/by-nc-nd/4.0/} to view a copy of this license. For any use beyond those covered by this license, obtain permission by emailing \href{mailto:info@vldb.org}{info@vldb.org}. Copyright is held by the owner/author(s). Publication rights licensed to the VLDB Endowment. \\
\raggedright Proceedings of the VLDB Endowment, Vol. \vldbvolume, No. \vldbissue\ %
ISSN 2150-8097. \\
\href{https://doi.org/\vldbdoi}{doi:\vldbdoi} \\
}\addtocounter{footnote}{-1}\endgroup

\ifdefempty{\vldbavailabilityurl}{}{
\vspace{.3cm}
\begingroup\small\noindent\raggedright\textbf{PVLDB Artifact Availability:}\\
The source code, data, and/or other artifacts have been made available at \url{\vldbavailabilityurl}.
\endgroup
}

\section{Introduction}
\begin{figure}
\centering
\includegraphics[width=\linewidth]{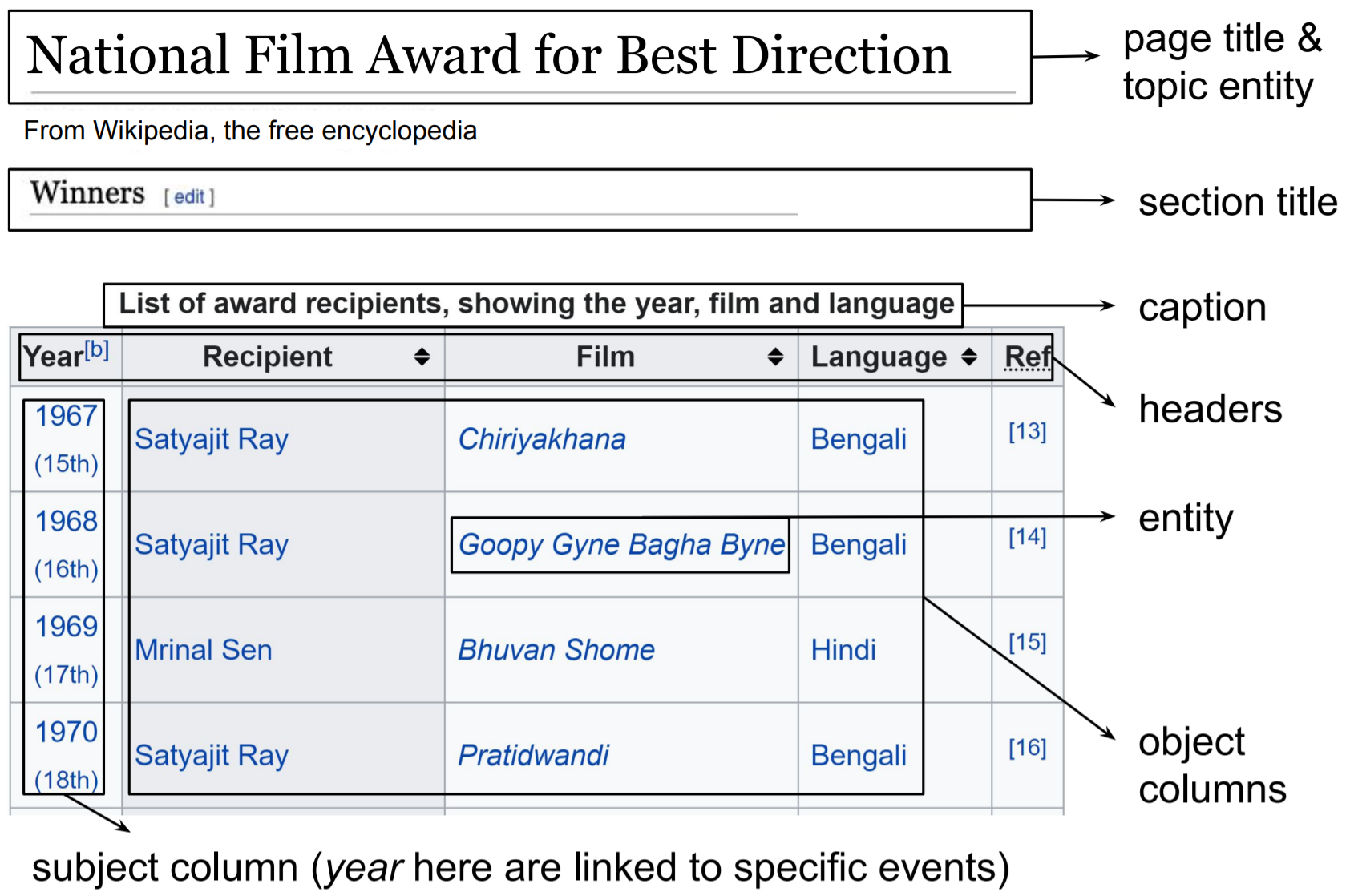}
\vspace{-15pt}
\caption{An example of a relational table from Wikipedia.}
\vspace{-10pt}
\label{fig:example}
\end{figure}
Relational tables are in abundance on the Web and store a large amount of knowledge, often with key entities in one column and attributes in the others.
\nop{\cy{relational tables vs semi-structured tables the two terms seem to be used interchangeably, pick one and stick to it, unless the difference is important, which then should be explained.}}
%Among them, relational tables (also referred to as entity-attribute tables) are a special class that stores a large amount of knowledge, usually with key entities as one core column and their attributes as other columns. 
Over the past decade, various large-scale collections of such tables have been aggregated \cite{cafarella2008webtables, CafarellaHZWW08, bhagavatula2015tabel, lehmberg2016large}. For example, \citet{cafarella2008webtables, CafarellaHZWW08} reported 154M relational tables out of a total of 14.1 billion tables in 2008. More recently, \citet{bhagavatula2015tabel} extracted 1.6M high-quality relational tables from Wikipedia. Owing to the wealth and utility of these datasets, various tasks such as table interpretation \cite{bhagavatula2015tabel,Zhang2020WebTE,Mulwad2010UsingLD,Ritze2015MatchingHT,Zhang2017EffectiveAE,Efthymiou2017MatchingWT}, table augmentation \cite{sarma2012find,zhang2017entitables,yakout2012infogather,ahmadov2015towards,zhang2019auto,cafarella2008webtables}, etc., have made tremendous progress in the past few years.  \nop{\citet{cafarella2008webtables, CafarellaHZWW08} first introduced the concept of relational tables and reported 154M such tables from a general web crawl in 2008. \citet{lehmberg2016large} used the Common Crawl corpus and extracted 233M web tables which are further classified into three categories: relational, entity, and matrix. \citet{bhagavatula2015tabel} focused on Wikipedia instead, and extracted 1.6M high quality relational tables.}

However, previous work such as \cite{zhang2017entitables,zhang2019auto,bhagavatula2015tabel} often rely on heavily-engineered task-specific methods such as  {simple statistical/language features or straightforward string matching}\nop{using simple language/statistical features for ranking and straightforward string matching for retrieval}. These techniques suffer from several disadvantages. First, simple features only capture shallow patterns and often fail to handle the flexible schema and varied expressions in Web tables. Second, task-specific features and model architectures require effort to design and do not generalize well across tasks.   %\hs{articulate disadvantages: (1) cannot capture the semantic meaning of different table components. (2) cannot generalize across different tasks.. (3) limited by the amount of task-specific supervision; does not utilize unsupervised data.} 

Recently, the pre-training/fine-tuning paradigm has achieved\linebreak\nop{inserted linebreak so that ``notable'' doesn't get broken into ``no-table'' which reads a little weird since we are dealing with tables.} notable success on unstructured text data. Advanced language models such as BERT \cite{devlin2019bert} can be pre-trained on large-scale unsupervised text and subsequently fine-tuned on downstream tasks using task-specific supervision. In contrast, little effort has been extended to the study of such paradigms on relational tables. Our work fills this research gap. %focused on developing task specific models

\nop{In particular, we focus on \textit{entity-focused relational tables}, which contain high-quality relational knowledge about entities... More specifically,} 
%Unsupervised representation learning on texts~\cite{} and knowledge bases (KBs)~\cite{} has gained much attention recently. In contrast, little work has been conducted on representation learning on semi-structured tables. \hs{Note: there is a disconnection between ``pretraining'' in last paragraph and ``unsupervised representation learning'' here. Do we still need this paragraph?} 

Promising results in some table based tasks were achieved by the representation learning model of \cite{Deng2019Table2VecNW}. This work serializes a table into a sequence of words and entities (similar to text data) and learns embedding vectors for words and entities using Word2Vec~\cite{mikolov2013distributed}. However, \cite{Deng2019Table2VecNW} cannot generate \textit{contextualized} representations, i.e., {it does not consider varied use of words/entities in different contexts and only produces a single fixed embedding vector for word/entity.} In addition, \textit{shallow} neural models like Word2Vec have relatively limited learning capabilities, which hinder the capture of complex semantic knowledge contained in relational tables. 
%that captures the semantic pattern in table schema as well as the knowledge embedded in table body.
\nop{we propose XXX, a novel table pre-training paradigm that learns representation for different elements of relational tables and benefits various table-driven tasks.}

\nop{~\\
 \xd{We can view the masked token prediction in BERT as recover based on the context (unmasked tokens), and the context can be on the left/right side, this is what they call it bi-directional. In table, actually there is no direction. We use the same idea of recover based on the context, here the context is defined by table structure (e.g. for an entity in table cell, the context is its column header, column in the same row/column and table caption)}\hs{In Transformer, you have bi-directional encoders, right? Let's clarify this somewhere. We probably don't need this detail in here, though.} \add{--TO REMOVE THIS PARAGRAPH}
~\\}

%\hs{Challenges \& How you would like to solve the problem:} 
We propose \modelnospace, a novel framework for learning deep contextualized representations on relational tables via pre-training in an unsupervised manner and task-specific fine-tuning. 

There are two main challenges in the development of \modelnospace: (1) \textit{Relational table encoding}. Existing neural network encoders are designed for linearized sequence input and are a good fit with unstructured texts. However, data in relational tables is organized in a semi-structured format. Moreover, a relational table contains multiple components including the table caption, headers and cell values. The challenge is to develop a means of modeling the row-and-column structure as well as integrating the heterogeneous information from different components of the table. (2) \textit{Factual knowledge modeling}. Pre-trained language models like BERT~\cite{devlin2019bert} and ELMo~\cite{Peters2018DeepCW} focus on modeling the syntactic and semantic characteristics of word use in natural sentences. However, relational tables contain a vast amount of factual knowledge\nop{\hs{In this paper, we focus on relations among entities and leave other types of knowledge such as numerical attributes as future work.}} about entities, which cannot be captured by existing language models directly. Effectively modelling such knowledge in \model is a second challenge.  %\hs{Do we have other challenges? These two, especially the first one, do not sound very challenging.}

To address the first challenge, we encode information from different table components into separate input embeddings and fuse them together. We then employ a \textit{structure-aware} Transformer \cite{vaswani2017attention} encoder with masked self-attention. The conventional Transformer model is a bi-directional encoder, so each element (i.e., token/entity) can attend to all other elements in the sequence. We explicitly model the row-and-column structure by restraining each element to only aggregate information from other structurally related elements. To achieve this, we build a visibility matrix based on the table structure and use it as an additional mask for the self-attention layer. %\hs{this ``so that'' is not clear. ``element'' or ``component'' - let's be consistent} \xd{the element here is different from the component above. Component is caption/header/content, while the element is a token/entity in the table }. 

For the second challenge, {we first learn embeddings for each entity during pre-training. We then model the relation between entities in the same row or column with the assistance of the visibility matrix. Finally, we propose a Masked Entity Recovery (MER) pre-training objective. The technique randomly masks out entities in a table with the objective of recovering the masked items based on other entities and the table context (e.g., caption/header). \textit{This encourages the model to learn factual knowledge from the tables and encode it into entity embeddings.} In addition, we utilize the entity mention by keeping it as additional information for a certain percentage of masked entities. \textit{This helps our model build connections between words and entities}} \nop{we introduce an auxiliary task that recovers the entity with extra entity mention information}\nop{do we have a name for this task separately, or it is part of MER?}. We also adopt the Masked Language Model (MLM) objective from BERT, which aims to model the complex characteristics of word use in table metadata.
% : (1) Masked language model (MLM), which aims to model the complex characteristic of word use in table metadata and diverse combination of table schema; (2) Masked entity recovery (MER), which intends to encode the factual knowledge embedded in Web tables and build connections between words and entities.  %\hs{Xiang, can you enrich this paragraph a bit? Check out the paragraphs above for the flow and logic.}
%\nop{let the model simultaneously learn the semantic pattern in table schema as well as the knowledge embedded in table body.}

\nop{We conduct experiments on several table-driven tasks, i.e., row population, cell filling and attribute discovery. Experimental results show that the representation learned by our pre-trained model significantly boosts the performance of baseline models in all tasks.}

We pre-train our model on ~570K relational tables from Wikipedia to generate contextualized representations for tokens and entities in the relational tables. We then fine-tune our model for specific downstream tasks using task-specific labeled data. A distinguishing feature of \model is its {universal}\nop{unified} architecture across different tasks - {only} minimal modification is needed to cope with each downstream task. To facilitate research in this direction, we {compiled a benchmark}\nop{\hs{minor: should we remove 'table understanding' as the last three seems different from the first three; call it 'a benchmark' or 'a benchmark for table understanding and augmentation' if needed?} \xd{Yes, previously we call the benchmark as table understanding, and further split the tasks into table interperation and table augmentation. But table understanding seems to be the same as interpretation.}} that consists of 6 diverse tasks, including entity linking, column type annotation, relation extraction, row population, cell filling and schema augmentation. \revision{}{We created new datasets in addition to including results for existing datasets when publicly available.} \nop{We used the dataset from \cite{Efthymiou2017MatchingWT} in TUBE for the entity linking task and create new datasets for the other tasks.} Experimental results show that \model substantially outperforms existing task-specific and shallow Word2Vec based methods. %For tasks where direct comparisons are possible, \model outperforms Table2Vec~\cite{Deng2019Table2VecNW}, which learns word and entity embeddings using Word2Vec as mentioned earlier. 

Our contributions are summarized as follows:

\begin{itemize}
    \item To the best of our knowledge, \model is the first framework that introduces the pre-training/fine-tuning paradigm to relational Web tables\nop{ \revision{}{for table understanding}}. The pre-trained representations along with the universal model design save tremendous effort on engineering task-specific features and architectures.
    \item We propose a structure-aware Transformer encoder to model the structure information in relational tables. \nop{We also present two novel pre-training objectives to learn the semantic, structure and knowledge information \hs{this part is unclear. Do you mean the two enity-related objectives?} in relational tables in an unsupervised manner with large-scale unlabeled data.} We also present a novel Masked Entity Recovery (MER) pre-training objective to learn the semantics as well as the factual knowledge about entities in relational tables.
    \item To facilitate research in this direction, we present \nop{TUBE, }a benchmark that consists of 6 different tasks for \revision{}{table interpretation and augmentation}. We show that \model generalizes well to various tasks and substantially outperforms existing models. Our source code, benchmark, as well as pre-trained models will be available online\nop{\hs{URL to code?}}.
\end{itemize}

%covering both table interpretation and table augmentation.

\begin{table}[t]
\centering
\caption{Summary of notations for our table data.}
\vspace{-10pt}
\begin{tabular}{cl}
\hline
    Symbol & Description \\
    \hline
    $T$   & A relational table $T=(C,H,E,e_t)$ \\
    $C$   & Table caption (a sequence of tokens) \nop{$C=\{w_0,w_1,w_2,...,w_n\}$} \\
    $H$   & Table schema $H=\{h_0,...,h_i,...,h_m\}$ \\
    $h_i$ & A column header (a sequence of tokens) \nop{$h_i=\{u_0,u_1,...,u_n\}$} \\
    $E$   & Columns in table that contains entities \\
    $e_t$ & The topic entity of the table $e_t = (e_t^\text{e},e_t^\text{m})$\\
    $e$ & An entity cell $e = (e^\text{e},e^\text{m})$\\
    \hline
\end{tabular}
\vspace{-15pt}
\label{tab:symbtable}
\end{table}
\section{Preliminary}
%\noindent\textbf{Relational table} 
We now present our data model and give a formal task definition: 

In this work, we focus on relational Web tables and are most interested in the factual knowledge about entities. Each table $T \in \mathcal{T}$ is associated with the following: (1) Table caption $C$, which is a short text description summarizing what the table is about. {When the page title or section title of a table is available, we concatenate these with the table caption.} (2) Table headers $H$, which define the table schema; (3) Topic entity $e_t$, which describes what the table is about and is usually extracted from the table caption or page title; (4) Table cells $E$ containing entities. Each entity cell $e \in E$ contains a specific object with a unique identifier. For each cell, we define the entity as $e = (e^\text{e},e^\text{m})$, where  $e^\text{e}$ is the specific entity linked to the cell and $e^\text{m}$ is the entity mention (i.e., the text string).  

$(C,H,e_t)$ is also known as \textit{table metadata} and $E$ is the actual \textit{table content}. Notations used in the data model are summarized in Table~\ref{tab:symbtable}.
%\add{should we use $E$ to mean table cells rather than $T^C$?}
%\xd{So originally in this paragraph I first introduce components of general web tables, the $T^C$ can be any cells in the table. The formal definition below is for relational tables, and $E$ is entity cells. I will adjust definitions in experiment section.} 

\nop{For any web table $T \in \mathcal{T}$, we have the following: (1) the table caption $C$ that is a short text description summarizing what the table is about, (2) the table headers $H$ that defines the table schema, (3) the topic entity $e_t$ that the table is centered around, usually extracted from the table caption or page title, (4) the table cells $T^C$ that contains cell values. $(C,H,e_t)$ is also known as \textit{table metadata}, while $T^C$ is the actual \textit{table content}.
In this work, we focus on relational tables. As we are most interested in the factual knowledge about entities, we further narrow the focus of our study to entity cells. An entity cell in the context of this work is a table cell that represent a specific object with a unique identifier. The data model used in this work is formally defined as follows.}
\nop{
\begin{definition}
We consider a relational table $T_R=(C,H,E,e_t)$, where $E$ are columns in the table that contain entity cells.\textst{$E$ can be further divided into the subject column $E_s$ containing subject entities (or core entities), and object columns $E_o$ describing attributes of the subject entities.} For each entity cell $e \in E$, we have $e = (e^\text{e},e^\text{m})$. $e^\text{e}$ is the specific entity linked to the cell while $e^\text{m}$ is the entity mention (i.e., the text string) in the cell. 
\end{definition}
}
\nop{\ww{Are non-entity columns used at all?  If not, we can just mention that upfront and point readers to entity column specification in Section 5.  We don't need a separate symbol $E$ for it.  Along the same line, I don't see $E_s$ and $E_o$ in the rest of the paper, so maybe get rid of these symbols as well.} \xd{No, we only use entity columns.}}

% \begin{definition}
% A relational table contains a set of subject entities (or core entities), along with their attributes. It has the following components:\\
% \noindent\textit{Table caption} $C$ is a short text description summarizing what the table is about.\\
% \noindent\textit{Table headers} $H$ is a list of column names. Headers are typically in the first row in the table.\\
% \noindent\textit{Subject column} $E_s$ is the primary key of the table, which contains subject entities.\\
% \noindent\textit{Object column} $E_o$ is the remainder columns of the table, which contains attribute values.\\
% \noindent\textit{Topic entity} $e_t$ is the entity that the table is centered around. Oftentimes $e_t$ can be extracted from the table caption or page title.\\
% \end{definition}

Explicitly, we study the unsupervised representation learning on relational Web tables, which is defined as follows.
\begin{definition}
%$\mathcal{T_R}$
Given a relational Web table corpus, our representation learning task aims to learn in an unsupervised manner a task-agnostic {contextualized} vector representation for each token in all table captions $C$'s and headers $H$'s and for each entity (i.e., all entity cells $E$'s and topic entities $e_t$'s).
\end{definition}
%\nop{different table components $C,H,E,e_t$}

\nop{one vector per caption, per caption token, per schema, per header, per header token? let's be very clear.}

\nop{is the symbol table meant to be comprehensive? it seems to be missing a few, e.g., $T^C$.}

\nop{The same sequence representation is used for $C$ and $h_i$, which seems to imply they are related, while they are not.}

\nop{The same symbol $e$ is used to represent both an entity and superscript of a linked entity---maybe use simple text e and m for linked entity and entity mention superscripts?}
\section{Related Work}

%\subsection{Representation Learning on Text and Knowledge Base}
\noindent \textbf{Representation Learning.} The pre-training/fine-tuning paradigm has drawn tremendous attention in recent years. Extensive effort has been devoted to the development of unsupervised representation learning methods for both unstructured text and structured knowledge bases, which in turn can be utilized for a wide variety of downstream tasks via fine-tuning.

Earlier work, including Word2Vec~\cite{mikolov2013distributed} and GloVe~\cite{Pennington2014GloveGV}, pre-train distributed representations for words on large collections of documents. \nop{The pre-training aims to capture both syntactic and semantic information in the textual corpora and the resulting representations are widely used as input embeddings for various machine learning models. The methodology offers significant improvements over randomly initialized parameters and does not require feature engineering.} The resulting representations are widely used as input embeddings and offer significant improvements over randomly initialized parameters. However, pre-trained word embeddings suffer from word polysemy: they cannot model varied word use across linguistic contexts. This complexity motivated the development of contextualized word representations~\cite{Peters2018DeepCW,devlin2019bert,Yang2019XLNetGA} \nop{like ELMo~\cite{Peters2018DeepCW}, BERT~\cite{devlin2019bert} and XLNet~\cite{Yang2019XLNetGA}}. Instead of learning fixed embeddings per word, these works construct language models that learn the joint probabilities of sentences. 
\nop{Based on the difference in pre-training objectives, this body of work can be further divided into two categories: autoregressive language modeling that seek to estimate the probability distribution of a text corpus~\cite{Peters2018DeepCW,Yang2019XLNetGA}, and autoencoding language modeling that aims to reconstruct the original data from corrupted input~\cite{devlin2019bert}.}Such pre-trained language models have had huge success and yield state-of-the-art results on various NLP tasks~\cite{wang2018glue}.

Similarly, unsupervised representation learning has also been adopted in the space of structured data like knowledge bases (KB) \revision{}{ and databases}. Entities and relations in KB have been embedded into continuous vector spaces that still preserve the inherent structure of the KB~\cite{Wang2017KnowledgeGE}. These entity and relation embeddings are utilized by a variety of tasks, such as KB completion~\cite{Bordes2013TranslatingEF, Wang2014KnowledgeGE}, relation extraction~\cite{Riedel2013RelationEW,Weston2013ConnectingLA}, entity resolution~\cite{Glorot2013ASM}, etc. \revision{}{Similarly, \cite{fernandez2019termite} learned embeddings for heterogeneous data in databases and used it for data integration tasks.}

More recently, there has been a corpus of work incorporating knowledge information into pre-trained language models~\cite{zhang2019ernie,peters2019knowledge}. ERNIE~\cite{zhang2019ernie} injects knowledge base information into a pre-trained BERT model by utilizing pre-trained KB embeddings and a denoising entity autoencoder objective. \nop{KnowBERT~\cite{peters2019knowledge} includes a multitask training regime and jointly trains an entity linker with the language model.} The experimental results demonstrate that knowledge information is extremely helpful for tasks such as entity linking, entity typing and relation extraction.

Despite the success of representation learning on text and KB, \revision{there has been no study exploring contextualized representation learning on relational Web tables.}{few works have thoroughly explored contextualized representation learning on relational Web tables. Pre-trained language models are directly adopted in \cite{li2020deep} for entity matching. Two recent papers from the NLP community \cite{Herzig2020TAPASWS, Yin2020TaBERTPF} study pre-training on Web tables to assist in semantic parsing or question answering tasks on tables. \nop{this part does not read well; semantic? Also, just based on this sentence, it is not clear how our work distinguishes from theirs. Can we say their pre-training (of existing table encoders) is for specific tasks like X and Y?} \nop{like table question answering and semantic parsing. }} In this work, we introduce \modelnospace, a {new} methodology for learning deep contextualized representations for relational Web tables that preserve both semantic and knowledge information. \revision{}{In addition,  we conduct comprehensive experiments on a much wider range of table-related tasks}.  

%\subsection{Table Interpretation}
\vspace{1mm}
\noindent \textbf{Table Interpretation.} The Web stores large amounts of knowledge in relational tables. Table interpretation aims to uncover the semantic attributes of the data contained in relational tables, and transform this information into machine understandable knowledge. This task is usually accomplished with help from existing knowledge bases. In turn, the extracted knowledge can be used for KB construction and population. 

There are three main tasks for table interpretation: entity linking, column type annotation and relation extraction~\cite{bhagavatula2015tabel,Zhang2020WebTE}. Entity linking is the task of detecting and disambiguating specific entities mentioned in a table. Since relational tables are centered around entities, entity linking is a key step for table interpretation, and a fundamental component to many table-related tasks~\cite{Zhang2020WebTE}. \nop{\cite{LimayeGirija2010AnnotatingAS} first introduced the idea of factor graph based entity linking and combined five features including TF-IDF scores and compatibility between elements.}
\cite{bhagavatula2015tabel} employed a graphical model, and used a collective classification technique to optimize a global coherence score for a set of entities in a table. \cite{Ritze2015MatchingHT} presented the T2K framework, which is an iterative matching approach that combines both schema and entity matching. More recently, \cite{Efthymiou2017MatchingWT} introduced a hybrid method that combines both entity lookup and entity embeddings, which resulted in superior performance on various benchmarks.

Column type annotation and relation extraction both work with table columns. The former aims to annotate columns with KB types while the latter intends to use KB predicates to interpret relations between column pairs. Prior work has generally coupled these two tasks with entity linking~\cite{Mulwad2010UsingLD,Ritze2015MatchingHT,Zhang2017EffectiveAE}. After linking cells to entities, the types and relations associated with the entities in KB can then be used to annotate columns. In recent work, column annotation without entity linking has been explored ~\cite{Chen2018ColNetET,Chen2019LearningSA,Hulsebos2019SherlockAD}. These works modify text classification models to fit relational tables and have shown promising results. Moreover, relation extraction on web tables has also been studied for KB augmentation \cite{deng2019leveraging,cannaviccio2018leveraging,sekhavat2014knowledge}.

%\subsection{Table Augmentation}
\vspace{1mm}
\noindent \textbf{Table Augmentation.} Tables are a popular data format to organize and present relational information. Users often have to manually compose tables when gathering information. It is desirable to offer some intelligent assistance to the user, which motivates the study of table augmentation~\cite{zhang2017entitables}. Table augmentation refers to the task of expanding a seed query table with additional data. Specifically, for relational tables this can be divided into three sub-tasks: row population for retrieving entities for the subject column~\cite{sarma2012find,zhang2017entitables}, cell filling that fills the cell values for given subject entities~\cite{yakout2012infogather,ahmadov2015towards,zhang2019auto} and schema augmentation that recommends headers to complete the table schema~\cite{cafarella2008webtables,zhang2017entitables}. For row population tasks, \cite{sarma2012find} searches \nop{are conducted} for complement tables that are semantically related to seed entities and the top ranked tables are used for population. \cite{zhang2017entitables} further incorporates knowledge base information with a table corpus, and develops a generative probabilistic model to rank candidate entities with entity similarity features. For cell filling, \cite{yakout2012infogather} uses the query table to search for matching tables, and extracts attribute values from those tables. More recently, \cite{zhang2019auto} proposed the CellAutoComplete framework that makes use of a large table corpus and a knowledge base as data sources, and incorporates preprocessing, candidate value finding, and value ranking components. In terms of schema augmentation, \cite{cafarella2008webtables} tackles this problem by utilizing an attribute correlation statistics database (ACSDb) collected from a table corpus. \cite{zhang2017entitables} utilizes a similar approach to the row population techniques and ranks candidate headers with sets of features.

\vspace{1mm}
\noindent \revision{}{\textbf{Existing benchmarks.}
Several benchmarks have been proposed for table interpretation: (1) T2Dv2 \cite{lehmberg2016large} proposed in 2016 contains 779 tables from various websites. It contains 546 relational tables, with 25119 entity annotations, 237 table-to-class annotations and 618 attribute-to-property annotations. (2) Limaye et al. \cite{LimayeGirija2010AnnotatingAS} proposed a benchmark in 2010 which contains 296 tables from Wikipedia. It was used in \cite{Efthymiou2017MatchingWT} for entity linking\nop{with 5278 entity matches}, and was also used in \cite{Chen2019LearningSA} for column type annotation\nop{ with 114 columns annotated with 8 DBpedia types}. (3) Efthymiou et al. \cite{Efthymiou2017MatchingWT} created a benchmark (referred to as ``WikiGS'' in our experiments) that includes 485,096 tables from Wikipedia. WikiGS was originally used for entity linking with 4,453,329 entity matches. \cite{Chen2019LearningSA} further annotated a subset of it containing 620 entity columns with 31 DBpedia types and used it for column type annotation. (4) The recent SemTab 2019 \cite{jimenez2020semtab} challenge also aims at benchmarking systems that match tabular data to KBs, including three tasks, i.e., assigning a semantic type to an column, matching a cell to an entity, and assigning a property to the relationship between two columns. It used sampled tables from T2Dv2 \cite{lehmberg2016large} and WikiGS \cite{Efthymiou2017MatchingWT} in the first two rounds, and automatically generated tables in later rounds. \\
\indent In contrast to table interpretation, few benchmarks have been released for table augmentation. Zhang et al. \cite{zhang2017entitables} studied row population and schema augmentation with 2000 randomly sampled Wikipedia tables in total for validation and testing. \cite{zhang2019auto} curated a test collection with 200 columns containing 1000 cells from Wikipedia tables for evaluating cell filling. \\
\indent Although these benchmarks have been used in various recent studies, they still suffer from a few shortcomings: (1) They are typically small sets of\nop{Most existing datasets are quite small, with few} sampled tables with limited annotations. (2) SemTab 2019 contains a large number of instances; however, most of them are automatically generated and lack metadata/context of the Web tables\nop{\hs{by 'context,' do we mean `metadata?'}}. In this work, we compile a larger benchmark covering both table interpretation and table augmentation tasks. We also use some of these existing datasets\nop{as supplementary testing sets} for more comprehensive evaluation. By leveraging large-scale relational tables on Wikipedia and a curated KB, we ensure both the size and quality of our dataset.
} 
\section{Methodology}
\begin{figure}
    \includegraphics[width=0.9\linewidth]{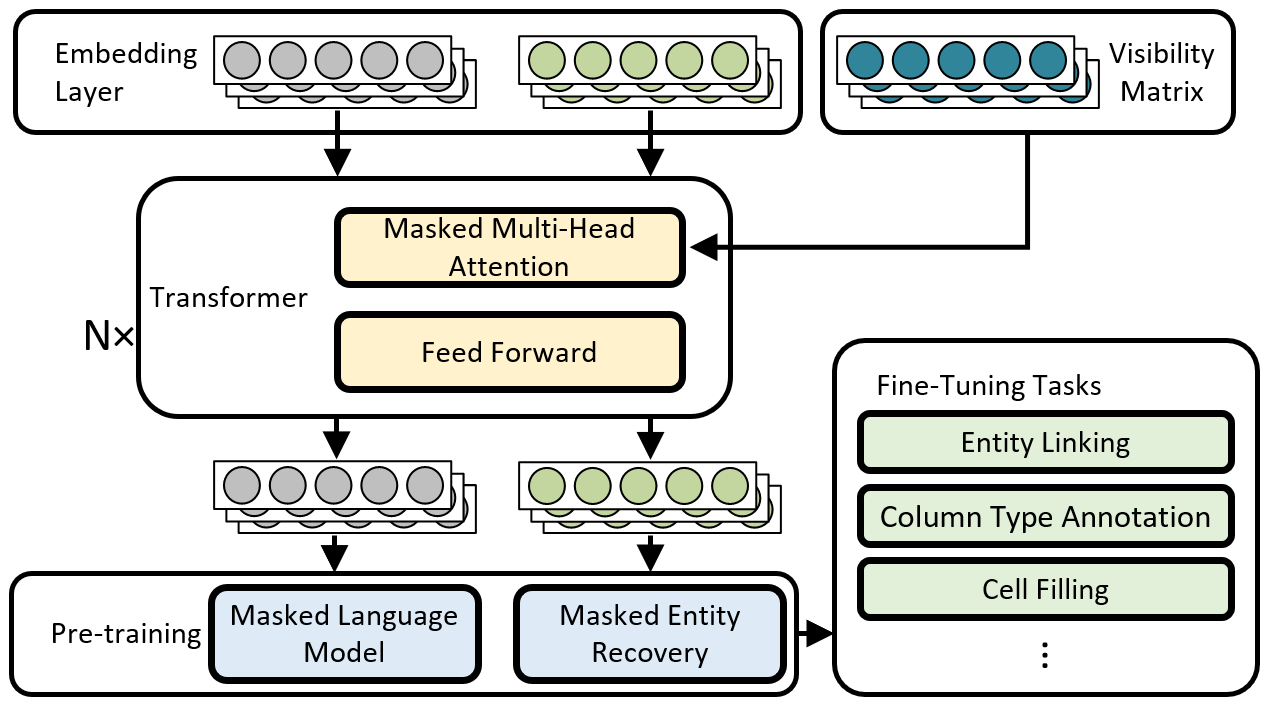}
    \centering
    \vspace{-10pt}
    \caption{Overview of our \model framework.
    }
    \vspace{-10pt}
    \label{fig:model}
\end{figure}
\begin{figure*}[t]
    \includegraphics[width=0.9\linewidth]{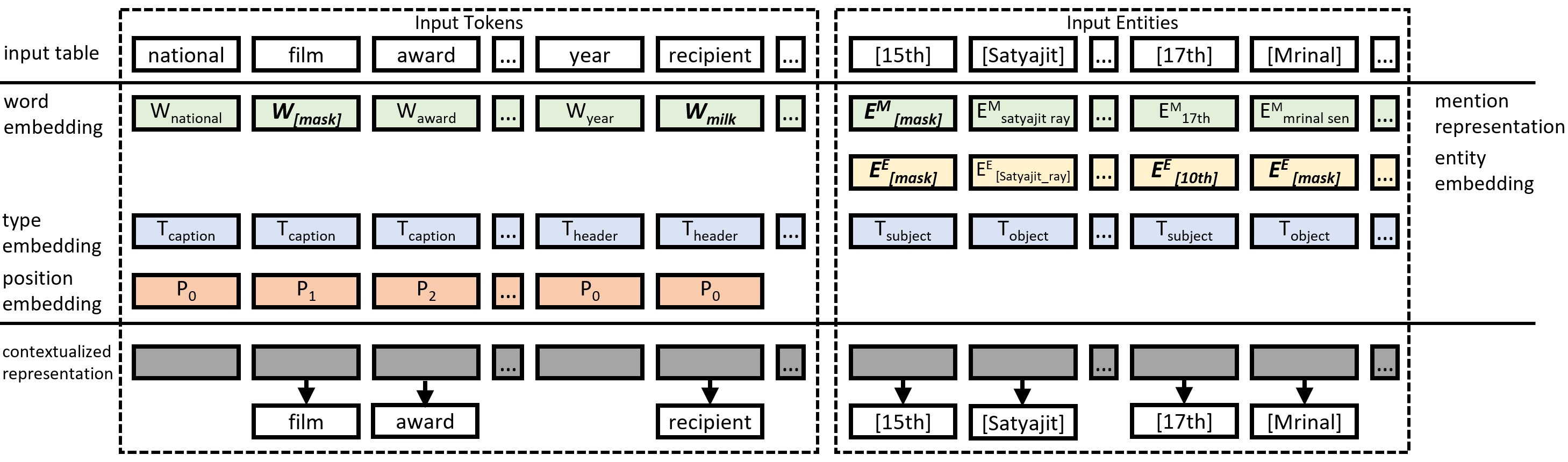}
    \vspace{-5pt}
    \centering
    \caption{Illustration of the model input-output. The input table is first transformed into a sequence of tokens and entity cells, and processed for structure-aware Transformer encoder as described in Section \ref{sec:pre-training}. We then get contextualized representations for the table and use them for pre-training. {Here [{\small \sf 15th}] (which means {\small \sf 15th National Film Awards}), [{\small \sf Satyajit}], ... are linked entity cells.}
    }
    \vspace{-5pt}
    \label{fig:input}
\end{figure*}
In this section, we introduce our \model framework for unsupervised representation learning on relational tables. \model is first trained on an unlabeled relational Web table corpus with  pre-training objectives carefully designed to learn word semantics as well as relational knowledge between entities. The model architecture is general and can be applied to a wide range of downstream tasks with minimal modifications. Moreover, the pre-training process alleviates the need for large-scale labeled data for each downstream task.

 %can adopt to new tasks with few fine-tuning steps on limited labeled data

\nop{and few finetuning steps on task-specific labeled data.}

\nop{worth mentioning that with the pre-trained model, there is less need for large amount of labeled data.}

\subsection{Model Architecture}
Figure \ref{fig:model} presents an overview of \model which consists of three modules: (1) an embedding layer to convert different components of an input table into input embeddings, (2) N stacked structure-aware Transformer \cite{vaswani2017attention} encoders to capture the textual information and relational knowledge, and (3) a final projection layer for pre-training objectives. {Figure \ref{fig:input} shows an input-output example.}

\subsection{Embedding Layer}
Given a table $T$$=$$(C,H,E,e_t)$, we first linearize the input into a sequence of tokens and entity cells by concatenating the table metadata and scanning the table content row by row. The embedding layer then converts each token in $C$ and $H$ and each entity in $E$ and $e_t$ into an embedding representation. %that can be fed into the structure-aware transformer encoder

%An example is given in Figure \ref{fig:input}.

\noindent\textbf{Input token representation.} \nop{We first concatenate $(C,H)$ as a sequence of tokens} For each token $w$, its vector representation is obtained as follows:
\nop{\ww{Maybe use a different index $j$ here.  A similar symbol $h_i$ has previously been used to denote the entirety of header $i$, which could cause confusion here.}\hs{check the current notations.}\ww{looks good.}}
\begin{equation}
    \mathbf{x}^\text{t}=\mathbf{w}+\mathbf{t}+\mathbf{p}.
\end{equation}

%\hs{can we change h to x? as h looks like hidden representation.} 
Here $\mathbf{w}$ is the word embedding vector, $\mathbf{t}$ is called the type embedding vector and aims to differentiate whether token $w$ is in the table caption or a header, and $\mathbf{p}$ is the position embedding vector that provides relative position information for a token within the caption or a header. \nop{for header, does the position capture which header it is within the schema, or just the token within the header?}\nop{Just the position within the header. We only differentiate between subject and object columns but do not use the column order} \nop{also regarding the position embedding, when you concatenate page title, section title, and caption into a single caption sequence, is the position of a (e.g., section title) token relative to the beginning of the concatenated sequence or the original sub-sequence?}\nop{to the beginning of the concatenated sequence.}

\noindent\textbf{Input entity representation.} For each entity cell $e = (e^\text{e}, e^\text{m})$ ({ same for topic entity $e_t$\nop{or topic entity $e_t = (e_t^\text{e},e_t^\text{m})$}}), we fuse the information from the linked entity $e^\text{e}$\nop{\add{($e_t^\text{e}$)}} and entity mention $e^\text{m}$\nop{ \add{($e_t^\text{m}$)}} together, and use an additional type embedding vector $\mathbf{t}^\text{e}$ to differentiate three types of entity cells (i.e., subject/object/topic entities). Specifically, we calculate the input entity representation $\mathbf{x}^\text{e}$ as:
%\cy{be consistent, previously you have $e^e$ first}
\begin{align}
    \mathbf{x}^\text{e}&=\mathtt{LINEAR}([\mathbf{e^\text{e}};\mathbf{e}^\text{m}])+\mathbf{t}^\text{e};\label{eq:entity_embedding}\\
    \mathbf{e}^\text{m}&=\mathtt{MEAN}({\mathbf{w}_{1},\mathbf{w}_{2},\dots, \mathbf{w}_{j}, \dots}).
\end{align}

\nop{Could you help clarify the above equations? Is the mean of words embeddings referring to the overall table topic entity $e_t$ mentioned in the preliminary section? If so, i am confused by the index *i*. Each word  in the header and caption and in cell value $ T^c$ has an individual token and entity embedding, but all of the entity embeddings $e_i^m$  are the same for the table independent of i? }\nop{So the entity here can be the topic entity (extracted from page title), or an entity in table cell. For each entity, we incorporate two sides of information: (1) the entity embedding for this specific entity learned by our model, (2) the surface form information. The $w_0,w_1,\dots,w_n$ here is the entity mention for each entity. So for an entity in a table cell, the $(w_i,\dots)$ here will be tokens in that cell. For topic entity, the $(w_i,\dots)$ will be the page title.}

Here $\mathbf{e}^\text{e}$ is the entity embedding learned during pre-training. To represent entity mention $e^\text{m}$, we use its average word embedding $\mathbf{w}_{j}$'s. \texttt{LINEAR} is a linear layer to fuse $\mathbf{e}^\text{e}$ and $\mathbf{e}^\text{m}$. 

{A sequence of token and entity representations ($\mathbf{x}^\text{t}$'s and $\mathbf{x}^\text{e}$'s) are then fed into the next module of \modelnospace, a structure-aware Transformer encoder, which will produce contextualized representations.}
\vspace{-15pt}

\nop{why do we need the subscripts $i$ and $j$ here?}\nop{removed them. changed the input embeddings to x's; made superscripts as text t and e; only vectors are bold and supercripts are not.}

% \begin{figure*}[t]
% \centering
% \includegraphics[width=0.9\linewidth]{figures/transformer.png}
% \vspace{-5pt}
% \caption{{Our Transformer-based structure-aware encoder.} \nop{figure is a bit blurry. to change later. you might want to highlight the visibility matrix, e.g., by using a different frame color.} {We propose a visibility matrix and use it as mask in scaled dot-product attention (See $M$ in Eqn. \ref{eq:att}) to model the structural information in a table.} \nop{this part is confusing: as mask in scaled dot-production attention}}
% \vspace{-0pt}
% \label{fig:transformer}
% \end{figure*}
\begin{figure}
    \centering
    \includegraphics[width=0.7\linewidth]{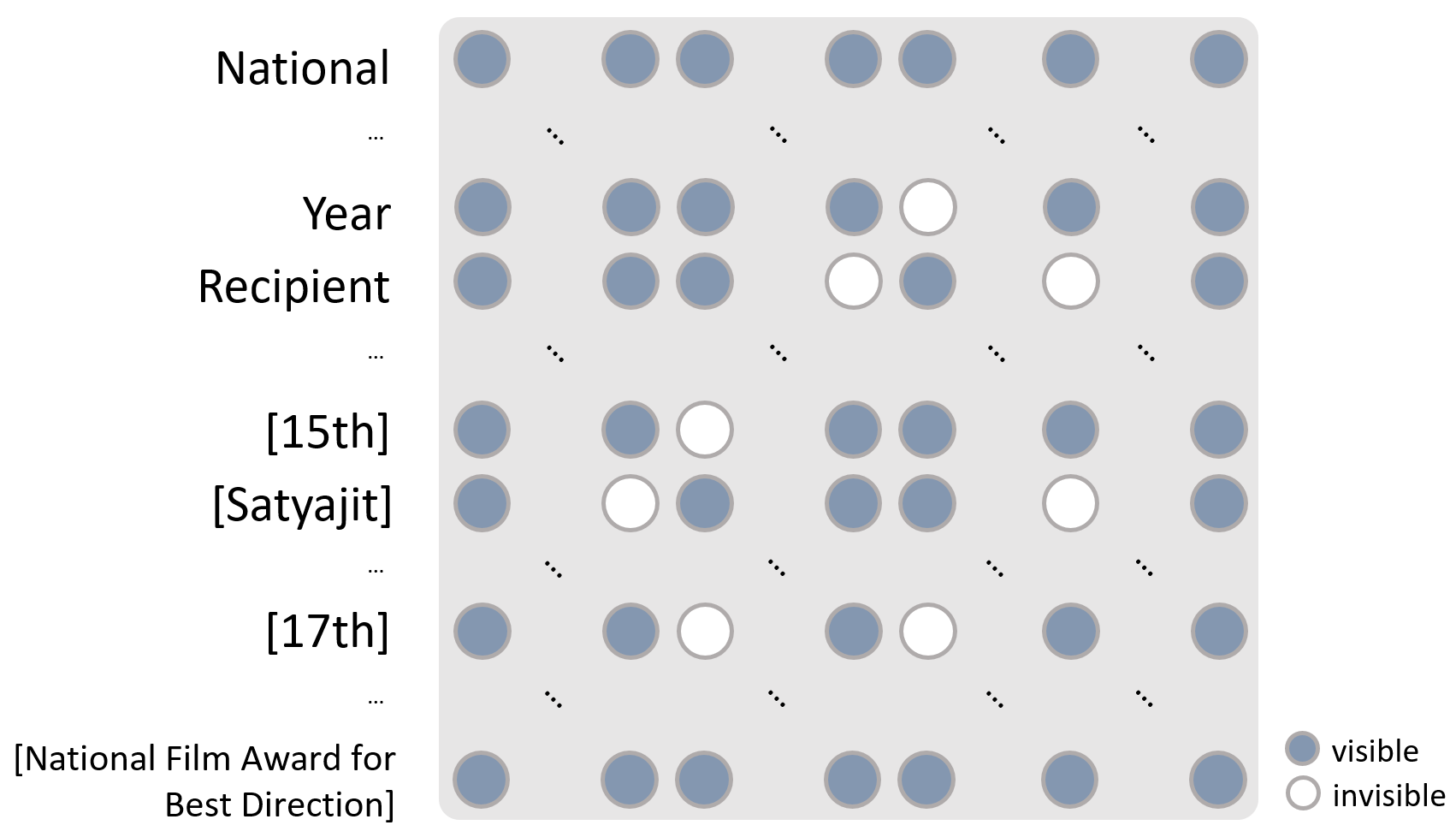}
    \vspace{-10pt}
    \caption{Graphical illustration of visibility matrix (symmetric).}
    \vspace{-5pt}
    \label{fig:vis_matrix}
\end{figure}
\begin{figure}
    \centering
    \includegraphics[width=0.90\linewidth]{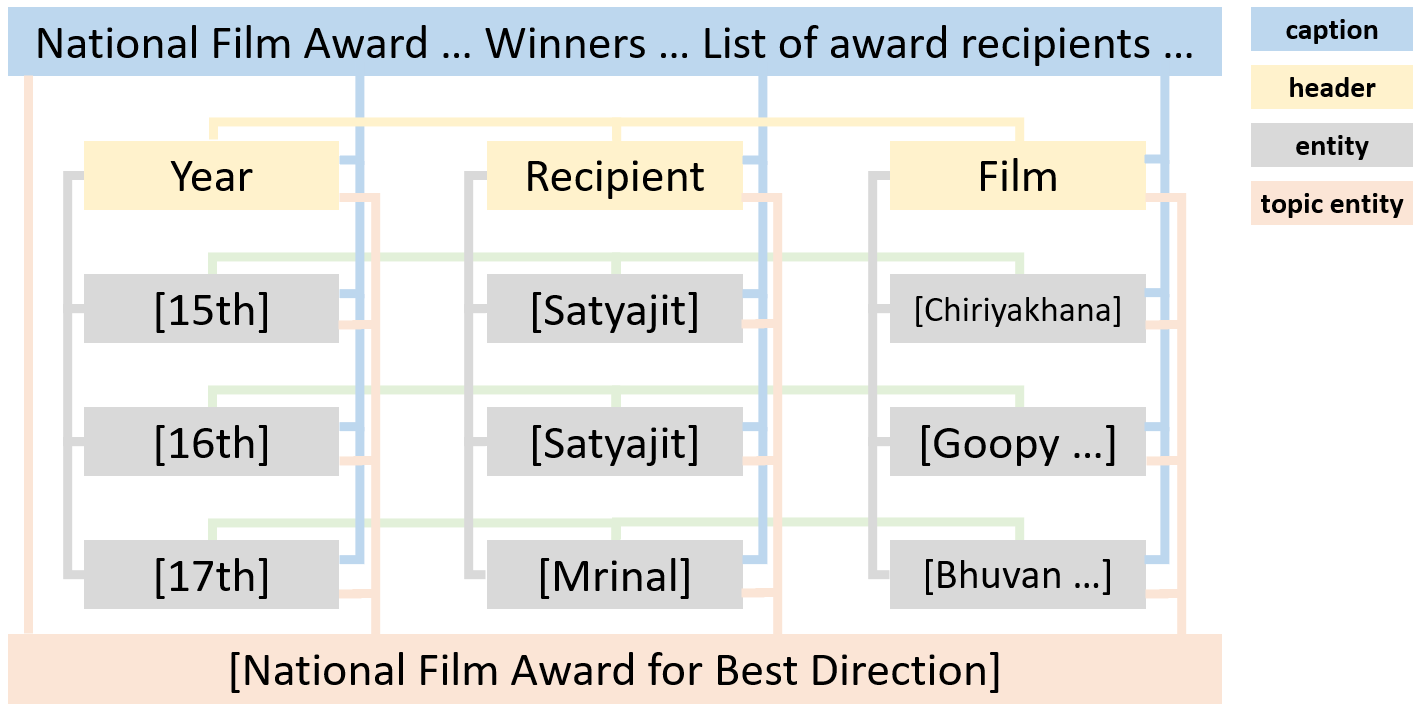}
    \vspace{-5pt}
    \caption{Graphical illustration of masked self-attention {by our visibility matrix}. Each token/entity in a table can only attend to its directly connected neighbors {(shown as edges here)}\nop{ strengthen the edges?}. \nop{See Example 4.1 for more explanation.}
    }
    \vspace{-15pt}
    \label{fig:mask_attention}
\end{figure}

\subsection{Structure-aware Transformer Encoder}

%\noindent\textbf{Transformer.} 
We choose Transformer \cite{vaswani2017attention} as our base encoder block, since it has been widely used in pre-trained language models \cite{devlin2019bert,radford2019language} and achieves superior performance on various natural language processing tasks~\cite{wang2018glue}. {Due to space constraints, we only briefly introduce the conventional Transformer encoder and refer readers to \cite{vaswani2017attention} for more details. Finally, we present a detailed explanation on our proposed visibility matrix for modeling table structure.}\nop{Since the usage of transformer has become the prevailing practice in NLP and data mining, we will omit the detailed description of the architecture and refer readers to \citep{vaswani2017attention}.} 

 \revision{As shown in Figure \ref{fig:transformer}, each}{Each} Transformer block is composed of a {multi-head} self-attention layer followed by a point-wise, fully connected layer \cite{vaswani2017attention}. Specifically, we calculate the multi-head attention as follows:
\begin{equation}
\begin{aligned}
\mathrm{ MultiHead }(\mathbf{h}) &=\mathrm{[head_{1};...;head_{i};...; head_{k}]} W^{O}; \\
\mathrm{head_i}&= \mathrm{Attention}\left(\mathbf{h} W_{i}^{Q}, \mathbf{h} W_{i}^{K}, \mathbf{h} W_{i}^{V}\right);\\
\mathrm{Attention}(Q, K, V)&=\mathrm{Softmax}\left(\frac{Q K^{T}}{\sqrt{d}}M\right) V.
\label{eq:att}
\end{aligned}
\end{equation}
Here $\mathbf{h}\in \mathbb{R}^{n\times d_\mathrm{model}}$ is the hidden state output from the previous Transformer layer or the input embedding layer and $n$ is the input sequence length. $\frac{1}{\sqrt{d}}$ \nop{removed subscript; originally $\frac{1}{\sqrt{d}}$} is the scaling factor. $W_{i}^{Q} \in \mathbb{R}^{d_{\mathrm{model}}\times d}$,$W_{i}^{K} \in \mathbb{R}^{d_{\mathrm{model}}\times d}$,$W_{i}^{V} \in \mathbb{R}^{d_{\mathrm{model}}\times d}$ and $W^{O} \in \mathbb{R}^{\mathrm{k}d\times d_\mathrm{intermediate}}$ \nop{there is no $W_{i}^{O}$, right?} are parameter matrices. For each head, we have $d=d_{\mathrm{model}}/\mathrm{k}$, where $\mathrm{k}$ is the number of attention heads. $M\in \mathbb{R}^{n\times n}$ is the visibility matrix which we detail next.

\noindent\textbf{Visibility matrix.} 
\nop{ NIT: Should it be called 'Visibility Matrix'?}
{To interpret relational tables and extract the knowledge embedded in them, it is important to model row-column structure.} For example, in Figure \ref{fig:example}, [\textsf{Satyajit}] and [\textsf{Chiriyakhana}] are related because they are in the same row, which implies that [\textsf{Satyajit}] directs [\textsf{Chiriyakhana}]. In contrast, [\textsf{Satyajit}] should not be related to [\textsf{Pratidwandi}]. Similarly, [\textsf{Hindi}] is a ``\textit{language}'' and its representation has little to do with the header ``\textit{Film}''.
We propose a visibility matrix $M$ to model such structure information in a relational table. Figure \ref{fig:vis_matrix} shows an example of $M$. 

Our visibility matrix acts as an attention mask so that each token (or entity) can only aggregate information from other structurally related {tokens/entities} during the self-attention calculation. $M$ is a symmetric binary matrix with $M_{i,j}=1$ if and only if $\mathtt{element}_j$ is visible to $\mathtt{element}_i$. The $\mathtt{element}$ here can be a token in the caption or a header, or an entity in a table cell.\nop{$M$ is a symmetric matrix.} Specifically, we define $M$ as follows:
    % \begin{itemize}
    %     \item If $\mathtt{element}_i$ or $\mathtt{element}_j$ is the topic entity or a token in table caption, $M_{i,j}=1$. \textit{Table caption and topic entity are visible to all components of the table.}
    %     \item If $\mathtt{element}_i$ is a token in a column header, $\mathtt{element}_j$ is also a token in a header or $\mathtt{element}_j$ is an entity in that column, $M_{i,j}=1$. \textit{Table headers are visible to each other, and to entities which belong to that column.}
    %     \item If $\mathtt{element}_i$ is an entity,  $\mathtt{element}_j$ is an entity in the same row or the same column, $M_{i,j}=1$. \textit{Table entities  in  the  same  row  or  same  column  are visible to each other.}
    % \end{itemize}

\nop{I think the visibility matrix overall can be defined more succinctly as the matrix representation of a symmetric, reflexive binary relation, where each instance into one of the two categories.}
\begin{itemize}
    \item If $\mathtt{element}_i$ is the topic entity or a token in table caption, $\forall j,M_{i,j}=1$. \textit{Table caption and topic entity are visible to all components of the table.}
    \item If $\mathtt{element}_i$ is a token or an entity in the table and $\mathtt{element}_j$ is a token or an entity in the same row or the same column, $M_{i,j}=1$.  \textit{Entities\nop{\st{(except for headers)}} and text content in the same row or the same column are visible to each other.}
\end{itemize}

\nop{The structure-aware transformer encoder can also be seen as Graph Neural Network (GNN) similar to \cite{Shaw2019GeneratingLF,Mller2019AnsweringCQ}. The input tokens and entities are nodes in the graph and edges are defined based on the table structure. \hs{is this paragraph necessary? move to related work if you want to discuss the connection? otherwise, people might ask why you don't use GNN? kind of invite questions..}}

\begin{example}
Use Figure \ref{fig:mask_attention} as an example. ``\textit{National Film ... recipients ...}'' \nop{changed to ``'' and italic to denote texts.} are tokens from the caption and [\textsf{National Film Award for Best Direction}] is the topic entity, and hence they can aggregate information from all other \texttt{elements}. ``\textit{Year}'' is a token from a column header, and it can attend to all \texttt{elements} \textit{except for} entity cells not belonging to that column. [\textsf{Satyajit}] is an entity from a table cell, so it can only attend to the caption, topic entity, entity cells in the same row/column as well as the header of that column.
\end{example}
\subsection{Pre-training Objective}\label{sec:pre-training}
\nop{To learn a universal representation for different table components that can be easily applied to various tasks, we design two pre-training objectives: Masked Language Model (MLM) for table metadata and Masked Entity Recovery (MER) for table content.}
{In order to pre-train our model on an unlabeled table corpus, we adopt the Masked Language Model (MLM) objective from BERT to learn representations for tokens in table metadata and propose a Masked Entity Recovery (MER) objective to learn entity cell representations.}

\nop{BTW, did we use word/token interchangeably? if so, may stick to one.}

\noindent\textbf{Masked Language Model.} %Inspired by language models like BERT, 
{We adopt the same Masked Language Model objective as BERT, which trains the model to capture the lexical, semantic and contextual information described by table metadata.} Given an input token sequence including table caption and table headers, we simply mask some percentage of the tokens at random, and then predict these masked tokens. \nop{it's a nice sentence, but I think we don't need it here as MLM isn't our contribution; we probably can shorten this entire MLM part: This is similar to denoising auto-encoders~\cite{vincent2008extracting}, but here we only predict the masked tokens rather than reconstructing the entire input.} {We adopt the same percentage settings as BERT.} The pre-training data processor selects 20\% of the token positions at random (note, we use a slightly larger ratio compared with 15\% in \cite{devlin2019bert} as we want to make the pre-training more challenging). \nop{\al{Why 20percent? BERT paper 15 percent.  Any difference?} \xd{The 20\% here is arbitrary. I don't tune the masking ratio for MLM, as the model seems does well on MLM task (given the accuracy of mask token prediction). I do tune the masking ratio for MER.}} \nop{If the i-th token is chosen,}{For a selected position,} (1) 80\% of the time we replace it with a special [MASK] token, (2) 10\% of the time we replace it with another random token, and (3) 10\% of the time we keep it unchanged. 

\begin{example}
Figure \ref{fig:input} shows an example of the above random process in MLM, where (1) ``\textit{film}'', ``\textit{award}'' and ``\textit{recipient}'' are chosen randomly, and (2) the input word embedding of ``\textit{film}'' is further chosen randomly to be replaced with the embedding of [MASK], (3) the input word embedding of ``\textit{recipient}'' to be replaced with the embedding of a random word ``\textit{milk}'', and (4) the input word embedding of ``\textit{award}'' to remain the same. \nop{check the text in headers; they should be italic.}
\end{example}
Given a token position selected for MLM, which has a contexutalized representation $\mathbf{h}^\text{t}$ output by our encoder, the probability of predicting its original token $w \in \mathcal{W}$ is then calculated as: \nop{check here; simplified.}
\begin{equation}
    P(w) = \frac{\mathrm{exp}\left(\mathtt{LINEAR}(\mathbf{h}^\text{t})\cdot\mathbf{w}\right)}{\sum_{w_k \in \mathcal{W}}{\mathrm{exp}\left(\mathtt{LINEAR}(\mathbf{h}^\text{t})\cdot\mathbf{w}_k\right)}}
\end{equation}

\vspace{1mm}
\noindent\textbf{Masked Entity Recovery.} In addition to MLM, we propose a novel Masked Entity Recovery (MER) objective to help the model capture the factual knowledge embedded in the table content as well as the associations between table metadata and table content. Essentially, we mask a certain percentage of input entity cells and then recover the linked entity based on surrounding entity cells and table metadata. This requires the model to be able to infer the relation between entities from table metadata and encode the knowledge in entity embeddings. 

{In addition, our proposed masking mechanism takes advantage of entity mentions.} Specifically, as shown in Eqn. \ref{eq:entity_embedding}, the input entity representation has two parts: the entity embedding $\mathbf{e}^\text{e}$ and the entity mention representation $\mathbf{e}^\text{m}$. For some percentage of masked entity cells, we only mask $\mathbf{e}^\text{e}$, and  as such the model receives additional entity mention information to help form predictions. This assists the model in building a connection between entity embeddings and entity mentions, and helps downstream tasks where only cell texts are available. 

\nop{Agree with with hs comment on why percentage picked.  Separately, i dont understand why 10 percent of the 20 percent in both tasks is left unchanged.  is this for evaluation?}
Specifically, \nop{The ratio for MER is picked by parameter search, but generally the model is not very sensitive to the it. We can show it in experiment. The unchanged part of input is used in BERT to mitigate the gap between pre-training and downstream task. Because when we actually apply the model, most input are correct.)} we propose the following masking mechanism for MER: The pre-training data processor chooses 60\% of entity cells \nop{should we call it entities or entity positions?} at random. Here we adopt a higher masking ratio for MER compared with MLM, because oftentimes in downstream tasks, none or few entities are given. For one chosen entity cell, (1) 10\% of the time we keep both $\mathbf{e}^\text{m}$ and $\mathbf{e}^\text{e}$ unchanged (2) 63\% (i.e., 70\% of the left 90\%) of the time we mask both $\mathbf{e}^\text{m}$ and $\mathbf{e}^\text{e}$ (3) 27\% (i.e., 30\% of the left 90\%) of the time we keep $\mathbf{e}^\text{m}$ unchanged, and mask $\mathbf{e}^\text{e}$ (among which we replace $\mathbf{e}^\text{e}$ with embedding of a random entity to inject noise in 10\% of the time\nop{do you mean, among the 27\%, 10\% is replaced with a random entity?}). {Similar to BERT, in both MLM and MER we keep a certain portion of the selected positions unchanged so that the model can generate good representations for non-masked tokens/entity cells. Trained with random tokens/entities replacing the original ones, the model is robust and utilizes contextual information to make predictions rather than simply copying the input representation.}
%\hs{you mean 10\% to random entity while 20\% to [mask]?} \xd{actually 10\% of the 27\%, so only around 2.7\% overall}

\begin{example}
Take Figure \ref{fig:input} as an example. \textsf{[15th]}, \textsf{[Satyajit]}, \textsf{[17th]} and \textsf{[Mrinal]} are first chosen for MER. Then, (1) the input mention representation and entity embedding of [\textsf{Satyajit}] remain the same. (2) The input mention representation and entity embedding of [\textsf{15th}] are both replaced with the embedding of [MASK] (3) The input entity embedding of [\textsf{Mrinal}] is replaced with embedding of [MASK], while the input entity embedding of [\textsf{17th}] is replaced with the embedding of a random entity [\textsf{10th}]. In both cases, the input mention representation are unchanged. \nop{what about their surfaces? no changes?} \nop{did we mention earlier that 10th/15th/.. are entities in case people confuse them with numbers?}
\end{example}

{Given an entity cell selected for MER with a contexutalized representation $\mathbf{h}^\text{e}$ output by our encoder, the probability of predicting entity $e \in \mathcal{E}$ is then calculated as follows. 
\begin{equation}
    P(e) = \frac{\mathrm{exp}\left(\mathtt{LINEAR}(\mathbf{h}^\text{e})\cdot\mathbf{e}^\text{e}\right)}{\sum_{e_k \in \mathcal{E}}{\mathrm{exp}\left(\mathtt{LINEAR}(\mathbf{h}^\text{e})\cdot\mathbf{e}_k^\text{e}\right)}}\label{eq:e_prob}
\end{equation}
\nop{here the vector e should be $e^\text{e}$?}
In reality, considering the entity vocabulary $\mathcal{E}$ is quite large, we only {use the above equation}\nop{use the system} to rank entities from a given candidate set. For efficient training, we construct the candidate set with (1) entities in the current table, (2) {entities that have co-occurred with those in the current table}, and (3) randomly sampled negative entities.}

 We use a cross-entropy loss function for both MLM and MER objectives and {the final pre-training loss is given as follows}:
 \begin{equation}
     loss = \sum{\mathrm{log}\left(P(w)\right)}+\sum{\mathrm{log}\left(P(e)\right)},
 \end{equation}
 {where the sums are over all tokens and entity cells selected in MLM and MER respectively.}
 \nop{I think we should make it more clear in terms of what the sum is over? over all selected token/entity positions in every table.}
 
 %\hs{an equation here?}
 
 ~\\
 \noindent \textbf{Pre-training details.}
In this work, we denote the number of Transformer blocks as N, the hidden dimension of input embeddings and all Transformer block outputs as $d_{\mathrm{model}}$, the hidden dimension of the fully connected layer in a Transformer block as $d_{\mathrm{intermediate}}$, and the number of self-attention heads as $\mathrm{k}$. {We take advantage of a pre-trained TinyBERT \cite{Jiao2019TinyBERTDB} model, which is a knowledge distilled version of BERT with a smaller size, and set the hyperparameters as follows:} \nop{In detail, we have the following model size:} N = 4, $d_{\mathrm{model}}=312$, $d_{\mathrm{intermediate}}=1200$, $\mathrm{k}=12$. 
{We initialize our structure-aware Transformer encoder parameters, word embeddings and position embeddings with TinyBERT \cite{Jiao2019TinyBERTDB}. \nop{To make entity embeddings aligned with other modules,} Entity embeddings are initialized using averaged word embeddings in entity names, and type embeddings are randomly initialized. We use the Adam \cite{Kingma2015AdamAM} optimizer with a linearly decreasing learning rate. The initial learning rate is 1e-4 chosen from [1e-3, 5e-4, 1e-4, 1e-5] {based on our validation set}. We pre-trained the model for 80 epochs.}
% \revision{}{Due to space constraints, we refer readers to the 
% extended version of our paper \cite{deng2020turl} for detailed ablation studies on the design choices. }
%\hs{maybe also mention training details like optimization algorithm, learning rate, etc. Not sure if we need to talk about ``complexity/parameter size/efficiency.''}
\section{Dataset Construction for Pre-training}
%\section{Dataset Construction for Pre-training}
We construct a dataset for unsupervised representation learning based on the WikiTable corpus \cite{bhagavatula2015tabel}, which originally contains around 1.65M tables extracted from Wikipedia pages. The corpus {contains a large amount of factual knowledge on} various topics ranging from sport events (e.g., Olympics) to artistic works (e.g., TV series). The following sections introduce our data construction process as well as characteristics of the dataset.
%\hs{a bit justification why this corpus is suitable to serve our purpose? e.g., why model pre-trained on it can help downstream tasks?}

\subsection{Data Pre-processing and Partitioning}
\label{sec:data}
\begin{table}[]
    \centering
    \caption{Dataset statistics (per table) in pre-training.}
    \vspace{-10pt}
    \resizebox{0.88\linewidth}{!}{\begin{tabularx}{0.95\linewidth}{|l|l|Y|Y|Y|Y|}
    \hline
    \multicolumn{1}{|c|}{} & split & \multicolumn{1}{c|}{min} & \multicolumn{1}{c|}{mean} & \multicolumn{1}{c|}{median} & \multicolumn{1}{c|}{max} \\ \hline
    \multirow{3}{*}{\# row} & train & 1 & 13 & 8 & 4670 \\
     & dev & 5 & 20 & 12 & 667 \\
     & test & 5 & 21 & 12 & 3143 \\ \hline
     \multirow{3}{*}{\# ent. columns} & train & 1 & 2 & 2 & 20 \\
     & dev & 3 & 4 & 3 & 15 \\
     & test & 3 & 4 & 3 & 15 \\ \hline
     \multirow{3}{*}{\# ent.} & train & 3 & 19 & 9 & 3911 \\
     & dev & 8 & 57 & 34 & 2132 \\
     & test & 8 & 60 & 34 & 9215 \\ \hline
    \end{tabularx}}
    \vspace{-10pt}
    \label{tab:data_stat}
\end{table}
\noindent\textbf{Pre-processing.} The corresponding Wikipedia page of a table often provides much contextual information, such as page title and section title that can aid in the understanding of a table topic. We concatenate page title, section title and table caption to obtain a comprehensive description.

In addition, each table in the corpus contains one or more header rows and several rows of  table content.\nop{Can any of them be empty?}\nop{table with empty components will get filtered out.} For tables with more than one header row, we concatenate headers in the same column to obtain one header for each column. For each cell, we obtain hyperlinks to Wikipedia pages in it and use them to normalize different entity mentions corresponding to the same entity. {We treat each Wikipedia page as an individual entity and do not use additional tools to perform entity linking with an external KB}. For cells containing multiple hyperlinks, we only keep the first link. We also discard {rows that have merged columns}\nop{merged rows} in a table\nop{what is a ``merged" row?}\nop{The column in this row is merged, e.g. some table about tv series, after row describing name/actor/director, there is one row contains storyline}. \nop{You mean rows in the middle of the table spanning (almost) all columns?}\nop{yes} \nop{We treat each Wikipedia page as an individual entity and do not perform entity linking with external KB.} %\hs{do we need to mention ``do not perform entity linking with external KB"?}.%We do not perform entity linking with external KB, but treat each Wikipedia page as individual entity.

\noindent\textbf{Identify relational tables.} We first locate all columns that contain at least one linked cell after pre-processing. We further filter out noisy columns with empty or illegal headers (e.g., \textit{note}, \textit{comment}, \textit{reference}, digit numbers\nop{single digit}\nop{why only single digit? won't numbers be filtered?}, etc.). The columns left are entity-centric and are referred to as entity columns. We then identify relational tables by finding tables that have a subject column. A simple heuristic is employed for subject column detection: the subject column must be located in the first two columns of the table and contain unique entities which we treat as subject entities. We further filter out tables containing less than three entities or more than twenty columns. {With this process, we obtain 670,171 relational tables.}

\noindent\textbf{Data partitioning.} From the above 670,171 tables, we select a high quality subset for evaluation: From tables that have (1) more than four linked entities in the subject column, (2) at least three entity columns including the subject column, and (3) more than half of the cells in entity columns are linked, we randomly select 10000 to form a held-out set. We further randomly partition this set into validation/test sets via a rough 1:1 ratio for model evaluation. All relational tables not in the evaluation set are used for pre-training. {In sum, we have 570171\,/\,5036\,/\,4964 tables respectively for pre-training/validation/test sets.}\nop{Just caught a mistake. So I used the randomsplit in spark to partition the dev/test with 0.5/0.5 ratio. I just noticed that the size of partition is approximate, so it's 5036/4964 instead of 5000 each.} %\hs{This ``evaluation'' is based on MLM and MER? not to be confused with downstream task evaluation.} \xd{The evaluation here includes pre-training and all downstream tasks. We always create test/val set from the held-out tables.}

\subsection{Dataset Statistics in Pre-training}
\nop{After pre-processing, we obtain 570,171 tables for training} \nop{this should appear above} Fine-grained statistics of our datasets are summarized in Table \ref{tab:data_stat}. {We can see that most tables in our pre-training dataset have moderate size, with median of 8 rows, 2 entity columns and 9 entities per table.}\nop{any comments/take-away msgs about the statistics?} We build a token vocabulary using the BERT-based tokenizer \cite{devlin2019bert} (with 30,522 tokens in total). For the entity vocabulary, we construct it based on the training table corpus and obtain 926,135 entities after removing those that appear only once. %\hs{in total or just the training set? could we also have \# of tables in Table \ref{tab:data_stat}?}

\begin{table}[t]
\centering
\caption{An overview of our benchmark tasks and strategies to fine-tune \modelnospace.}
\vspace{-10pt}
\resizebox{0.9\linewidth}{!}{\begin{tabular}{|c|>{\centering\arraybackslash} m{7cm}|}
\hline
\multicolumn{1}{|c|}{Task} & \multicolumn{1}{c|}{Finetune Strategy} \\ \hline
 \multirow{3}{*}{\rotatebox[origin=c]{90}{Table Interpretation\hspace{18pt}}}& \vspace{2pt}\includegraphics[width=\linewidth]{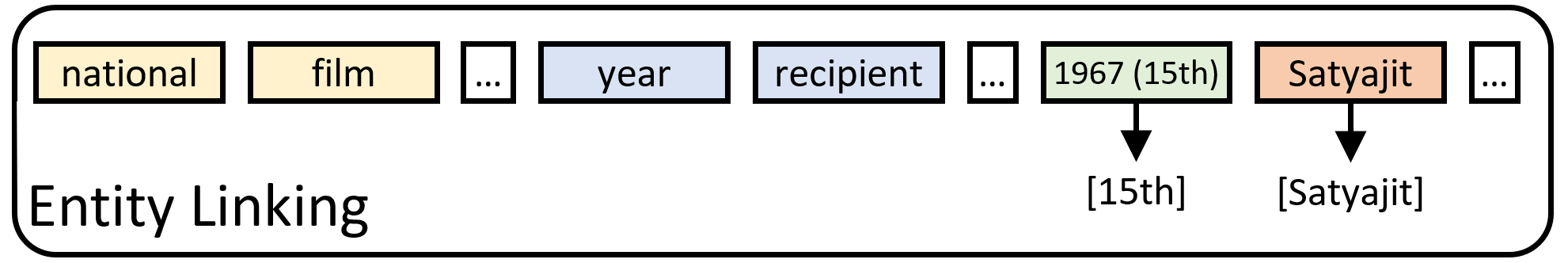} \\ \cline{2-2} 
 &\vspace{2pt}\includegraphics[width=\linewidth]{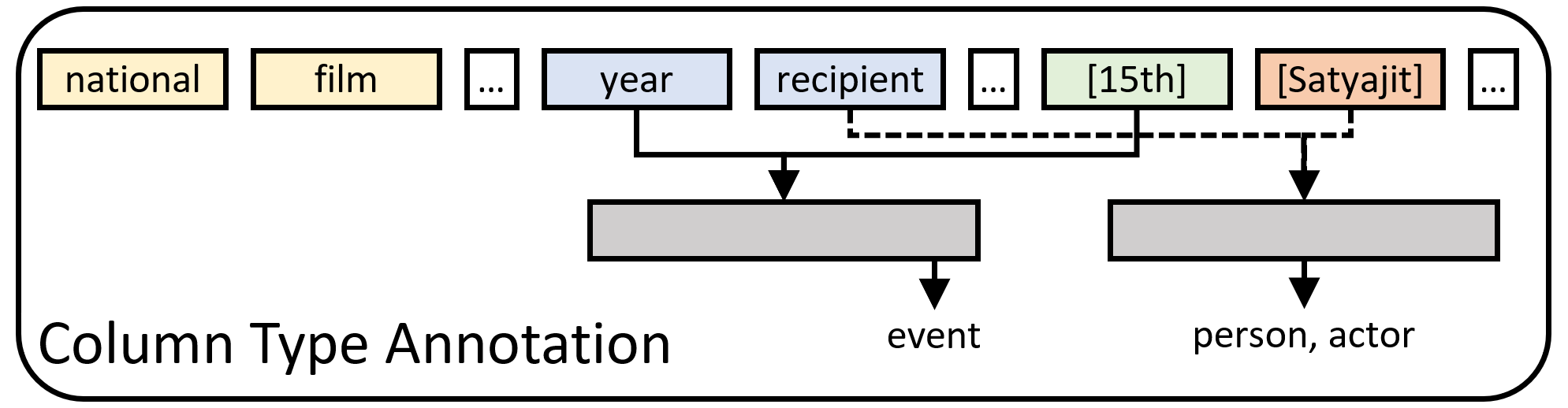}  \\ \cline{2-2} 
 & \vspace{2pt}\includegraphics[width=\linewidth]{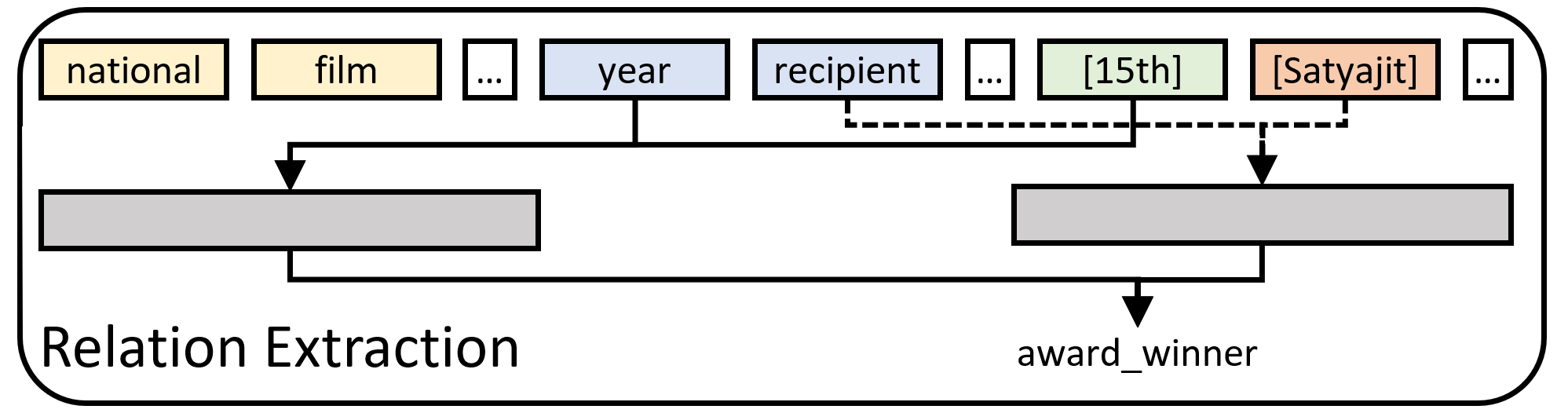}  \\ \hline
 \multirow{3}{*}{\rotatebox[origin=c]{90}{Table Augmentation\hspace{9pt}}}& \vspace{2pt}\includegraphics[width=0.9\linewidth]{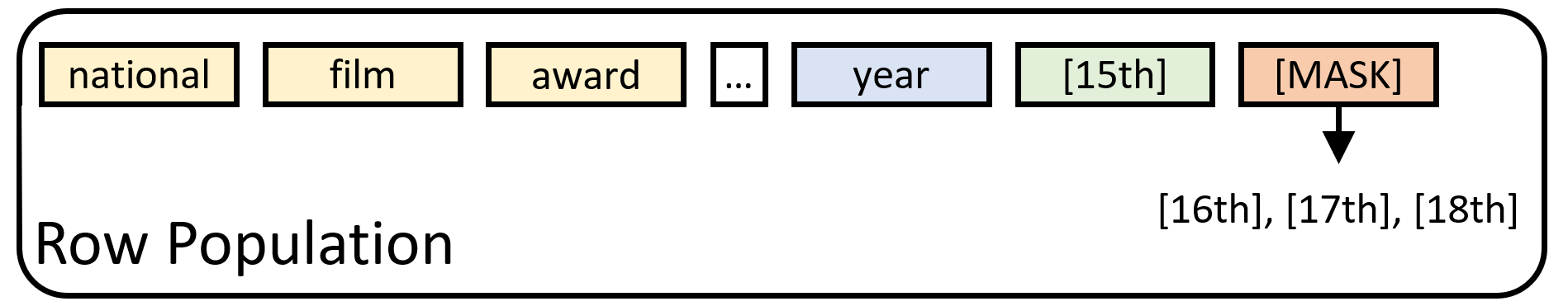} \\ \cline{2-2} 
 & \vspace{2pt}\includegraphics[width=\linewidth]{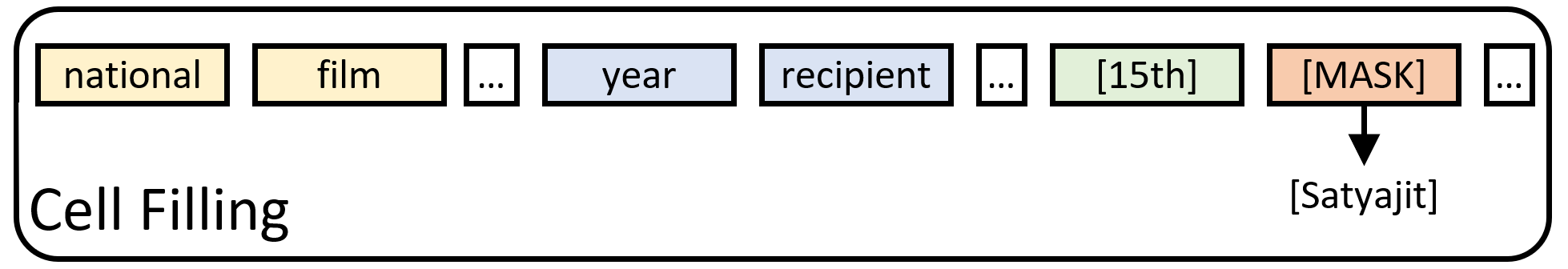}  \\ \cline{2-2} 
 & \vspace{2pt}\includegraphics[width=0.7\linewidth]{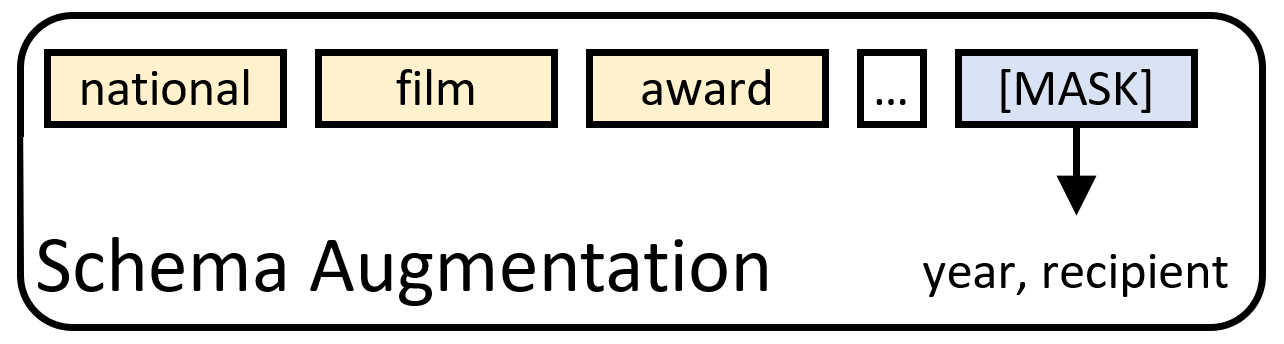}  \\ \hline
\end{tabular}}
\vspace{-10pt}
\label{tab:task_overview}
\end{table}
\section{Experiments}

% \begin{figure}
% \centering
% \includegraphics[width=0.55\textwidth]{figures/finetune.png}
% \caption{Illustration of fine-tuning on different tasks.}
% \label{fig:finetune_example}
% \end{figure}
To systematically evaluate our pre-trained framework as well as facilitate research, we compile a table understanding benchmark consisting of 6 widely studied tasks covering table interpretation (e.g., entity linking, column type annotation, relation extraction) and table augmentation (e.g., row population, cell filling, schema augmentation). {We include existing datasets for entity linking. However, due to the lack of large-scale open-sourced datasets, we create new datasets for other tasks based on our held-out set of relational tables and an existing KB.} %All our source code and the benchmark datasets will be available online\footnote{URL for code and data.}.  

Next we introduce the definition, baselines, dataset and results for each task. {Our pre-trained framework is general and can be fine-tuned for all the independent tasks.} %with fine-tuning. %with fine-tuning based on their corresponding datasets. 
\subsection{General Setup across All Tasks}

\nop{maybe use this section to talk about general important things for all tasks, e.g., what to be updated during fine-tuning, our dataset creation is based on partitions in pre-training and hence no overlap, fine-tuning epocs at 10}
{We use the pre-training tables to create the training set for each task, and always build data for evaluation using the held-out validation/test tables. This way we ensure that there is no overlapping tables in training and validation/test. For fine-tuning, we initialize the parameters with a pre-trained model, and further train all parameters with a task-specific objective. To demonstrate the efficiency of pre-training, we only fine-tune our model for 10 epochs unless otherwise stated.}

\subsection{Entity Linking}
\label{sec:EL}
Entity linking is a fundamental task in table interpretation, which is defined as:%as it is a key step to many other table related tasks. %The task can be stated as follows.
% \begin{definition}
% Given a table $T=(C, H, T^C)$ and a KB, for $T^C_{i,j}$ contains cell value $\mathbf{s}$, entity linking aims to identify the specific entity $\mathbf{e}$ this cell refers to in KB. \hs{do we need these symbols? can we simply say: Given a table $T$ and a KB, entity linking aims to identify which entity in the KB each cell value in $T$ refers to. }
% \end{definition}
\begin{definition}
Given a table $T$ and a knowledge base $\mathcal{KB}$, entity linking aims to \nop{\st{identify
and }}link each potential mention in cells of $T$ to its referent entity $e \in \mathcal{KB}$.
\end{definition}

Entity linking is usually addressed in two steps:\nop{it seems entity mention detection is also included in the definition?} a candidate generation module first proposes a set of potential entities, and an entity disambiguation module then ranks and selects the entity that best matches the surface form and is most consistent with the table context. Following existing work \cite{bhagavatula2015tabel,Ritze2015MatchingHT,Efthymiou2017MatchingWT}, we focus on entity disambiguation and use an existing Wikidata Lookup service\nop{\url{https://www.wikidata.org/w/api.php}} for candidate generation.

\vspace{1mm}
\noindent\textbf{Baselines.} We compare against the most recent methods for table entity linking T2K~\cite{Ritze2015MatchingHT}, Hybrid II~\cite{Efthymiou2017MatchingWT}, as well as the off-the-shelf Wikidata Lookup service. \nop{T2K uses an iterative process to jointly map columns to properties, rows to entities \add{(the cell in subject column of each row)} and tables to classes {(where the subject column type is treated as the table class)} until convergence.} {T2K uses an iterative matching approach that combines both schema and entity matching.} Hybrid II~\cite{Efthymiou2017MatchingWT} combines a lookup method with an entity embedding method. \nop{It first builds a disambiguation graph based on entity embeddings learned from a KB, and then applies PageRank algorithm~\cite{zwicklbauer2016doser} to select the set of entities that are most related to each other.} For Wikidata Lookup, we simply use the top-1 returned result as the prediction.

\vspace{1mm}
\noindent\textbf{Fine-tuning \modelnospace.} Entity disambiguation is\nop{essential for} essentially matching a table cell with candidate entities. We treat each cell as a potential entity, and input its cell text (entity mention $\mathbf{e}^\text{m}$ in Eqn. \ref{eq:entity_embedding}) as well as table metadata to our Transformer encoder and obtain a contextualized representation $\mathbf{h}^\text{e}$ for each cell. To represent each candidate entity, we utilize the name and description\nop{(returned by Wikidata Lookup)} as well as type information from a KB\nop{knowledge base}\nop{(e.g., DBpedia)}. The intuition is that when the candidate generation module proposes multiple entity candidates with similar names, we will utilize the description and type information\nop{in a KB} to find the candidate that is most consistent with the table context. Specifically, for a KB entity $e$, given its name $N$ and description $D$ (both are a sequence of words) and types $T$, we get its representation $\mathbf{e}^{\text{kb}}$ as follows: \nop{this seems not clear. do you mean, we first apply our model to the given table and get the representation of a cell value and then match it with the embedding of an entity? Is the entity embedding learnt by our pre-training as well?}
\nop{Adding to Huan's comment, if the cell representation does not include the initial entity embedding, can you specify what is meant here? i.e. i thought a cell was concatenation of word embedding + entities }\nop{we do not use the entity embedding learned in pre-training here, as we are doing entity linking so the cells are not linked. Also since we are using KB, we need to get representation for all entities in KB, not only those we seen in pre-training phase.}
\begin{equation}
    \mathbf{e}^{\text{kb}} = [\mathtt{MEAN}_{w\in N}\left(\mathbf{w}\right),\mathtt{MEAN}_{w\in D}\left(\mathbf{w}\right),\mathtt{MEAN}_{t\in T}\left(\mathbf{t}\right)].
    \label{eq:kb_repr}
\end{equation}
\nop{Check if these notations are more clear.} {Here, $\mathbf{w}$ is the embedding for word $w$,\nop{is $w$ frozen, sounds like it? if so, be explicit} which is shared with the embedding layer of pre-trained model. $\mathbf{t}$ is the embedding for entity type $t$ to be learned during this fine-tuning phase.} We then calculate a matching score between $\mathbf{e}^{\text{kb}}$ and $\mathbf{h}^\text{e}$ similarly as Eqn. \ref{eq:e_prob}. {We do not use the entity embeddings pre-trained by our model here, as the goal is to link mentions to entities in a target KB, not necessarily those appear in our pre-training table corpus.}\nop{we discussed this, but do we need to explain here why your learnt entity embeddings are not directly used for this task? mention there are likely unseen entities in evaluation?} The model is fine-tuned with a cross-entropy loss. \nop{did pre-trained parameters get updated like the above word embeddings?}

\vspace{1mm}
\noindent\textbf{Task-specific Datasets.} We use \revision{two}{three} datasets to compare different entity linking models: (1) We adopt the Wikipedia gold standards (WikiGS) dataset from \cite{Efthymiou2017MatchingWT}, which contains 4,453,329 entity mentions extracted from 485,096 Wikipedia tables and links them to DBpedia~\cite{Auer2007DBpediaAN}. (2) \nop{your rationale for creating this one, since they are also from Wikipedia..}{{Since tables in WikiGS also come from Wikipedia, some of the tables may have already been seen during pre-training, despite their entity linking information is mainly used to {pre-}train entity embeddings (which are not used here)}. For a better comparison, we also create our own test set from the held-out test tables mentioned in Section \ref{sec:data}, which contains 297,018 entity mentions from 4,964 tables. \revision{}{(3) To test our model on Web tables (i.e., those from websites other than Wikipedia), we also include the T2D dataset \cite{lehmberg2016large} which contains 26,124 entity mentions from 233 Web tables.\footnote{\revision{}{We use the data released by \cite{Efthymiou2017MatchingWT} (\url{https://doi.org/10.6084/m9.figshare.5229847}).}}} We use names and descriptions returned by Wikidata Lookup, and entity types from DBpedia.} %\hs{so, why not 5000?}\xd{changed, I originally put the number of mentions that have ground truth in the candidate set }. 

The training set for fine-tuning \model is based on our pre-training corpus, but with tables in the above WikiGS removed. We also remove duplicate entity mentions {and mentions where Wikidata Lookup fails to return the ground truth entity in candidates,} and finally obtain 1,264,217 entity mentions in 192,728 tables to fine-tune our model for the entity linking task. %We use the entity name and description returned by Wikidata Lookup, and obtain entity types in DBpedia. %\hs{so, we remove those tables that appear in our pre-training corpus from this task dataset? or, you change our pre-training corpus?} \hs{need more details about this dataset as the paper needs to be self-contained.} %the entity linking module.

\vspace{1mm}
\noindent\textbf{Results.} We set the maximum candidate size for Wikidata Lookup at 50 and also include the result of a Wikidata Lookup (Oracle), which considers an entity linking instance as correct if the ground-truth entity is in the candidate set. {Due to lack of open-sourced implementations, we directly use the results of T2K and Hybrid II in \cite{Efthymiou2017MatchingWT}.} {We use F1, precision (P) and recall (R) measures for evaluation. False positive is the number of mentions where the model links to wrong entities, not including the cases where the model makes no prediction (e.g., Wikidata Lookup returns empty candidate set).} %Performance  the entity linking results in .From the results we can see that 

As shown in Table \ref{tab:EL}\nop{ and \ref{tab:EL_own}}, our model gets the best F1 score and substantially improves precision \revision{}{on WikiGS and our own test set}\nop{we need to briefly mention earlier that we use these measures; i feel the measures are not clearly defined}. The disambiguation accuracy on WikiGS is 89.62\% (predict the correct entity if it is in the candidate set\nop{do need to introduce accuracy? or simply can see this from p/r/f1?}). A more advanced candidate generation module can help achieve better results in the future. We also conduct an ablation study on our model by removing the description or type information of a candidate entity {from Eqn. \ref{eq:kb_repr}}. From Table \ref{tab:EL}, we can see that entity description is very important for disambiguation, while entity type information only results in a minor improvement. This is perhaps due to the incompleteness of DBpedia, where a lot of entities have no types assigned or have missing types. \nop{do we have a comparison over others: In addition, our model does not conduct an iterative process, which greatly improve the scalability. On a single GPU, our model can process 431 tables per second }\\\indent \revision{}{On the T2D dataset, all models perform much better than on the two Wikipedia datasets, mainly because of its smaller size and limited types of entities. {The Wikidata Lookup baseline achieved high performance, and re-ranking using our model does not further improve. However, we adopt simple reweighting\footnote{\revision{}{We simply reweight the score of the top-1 prediction by our model with a factor of 0.8 and compare it with the top-1 prediction returned by Wikidata Lookup. The higher one is chosen as final prediction.}} to take into account the original result returned by Wikidata Lookup,}\nop{However, simple reranking using our model does not further improve over the Wikidata Lookup baseline. Given the high performance of Wikidata Lookup, we adopt simple reweighting to take into account the original order returned by the lookup service.} which brings the F1 score to 0.82. This demonstrates the potential of using features such as entity popularity (used in Wikidata Lookup) and ensembling strong base models. Additionally, we conduct an error analysis on T2D comparing our model (TURL + fine-tuning + reweighting) with Wikidata Lookup. From Table \ref{tab:EL_case}, \nop{I think the explanation here does not seem plausible. For the last two examples in table 5, we actually inferred the wrong types.} we can see that\nop{our model is good at inferring the type based on the context and rerank the candidate list accordingly. However,} while in many cases, our model can infer the correct entity type based on the context and re-rank the candidate list accordingly, it makes mistakes when there are entities in the KB that look very similar to the mentions.} \revision{}{To summarize, Table \ref{tab:EL} and \ref{tab:EL_case} show that there is room for further improvement of our model on  entity linking, which we leave as future work.}

\nop{General comment: Would be nice to include conclusion of model results versus baselines. What are advantages and disadvantages?}
\begin{table}[]
    \centering
    \caption{Model evaluation on entity linking task. \revision{The top half shows results on Wikipedia gold standards. The bottom half shows results on our testing set.}{All three datasets are evaluated with the same TURL + fine-tuning model.} \nop{In all the tables, if we directly copied numbers from a paper like T2K, maybe add a citation in the tables?}}
    \vspace{-10pt}
    \resizebox{0.9\linewidth}{!}{\begin{tabularx}{\linewidth}{|l|YYY|YYY|YYY|}
    \hline
    \multirow{2}{*}{Method} & \multicolumn{3}{c|}{WikiGS} & \multicolumn{3}{c|}{Our Test Set} & \multicolumn{3}{c|}{\revision{}{T2D}}\\
    \cline{2-10}
    & F1 & P & R & F1 & P & R & F1 & P & R\\
    \hline
    T2K \cite{Ritze2015MatchingHT} & 34 & 70 & 22 & - & - & - & 82 & \textbf{90} & 76\\
    Hybrid II \cite{Efthymiou2017MatchingWT} & 64 & 69 & \textbf{60} & - & - & - & \textbf{83} & 85 & \textbf{81}\\
    Wikidata Lookup & 57 & 67 & 49 & 62 & 62 & 60 & 80 & 86 & 75\\
    \hline
    TURL + fine-tuning & \textbf{67} & \textbf{79} & 58 & \textbf{68} & \textbf{71} & \textbf{66} & 78 & 83 & 73\\
    \hspace{10pt} w/o entity desc. & 60 & 70 & 52 & 60 & 63 & 58 & - & - & -\\
    \hspace{10pt} w/o entity type & 66 & 78 & 57 & 67 & 70 & 65 & - & - & -\\
    \hspace{10pt} + reweighting & - & - & - & - & - & - & 82 & 88 & 77\\
    \hline
    WikiLookup (Oracle) & 74 & 88 & 64 & 79 & 82 & 76 & 90 & 96 & 84\\
    \hline
    \end{tabularx}}
    \vspace{-10pt}
    \label{tab:EL}
\end{table}

\begin{table*}[]
    \centering
    \caption{Further analysis for entity linking on T2D corpus.}
    \vspace{-10pt}
    \resizebox{0.98\linewidth}{!}{\revision{}{\begin{tabular}{|c|c|c|c|c|c|}
    \hline
    Mention&Page title&Header&Wikidata Lookup result&TURL + fine-tuning + reweighting result&Improve\\
    \hline
    philip & \makecell{List of saints} & Saint&\makecell{Philip, male given name}&\makecell{Philip the Apostle, Christian saint\\ and apostle}&Yes\\
    \hline
    haycock & \makecell{All 214 Wainwright fells from\\ the pictorial guides - Wainwright Walks} &Fell Name&\makecell{Haycock, family name}&\makecell{Haycock, mountain in United Kingdom}&Yes\\
    \hline
    \makecell{don't you forget\\ about me} & \makecell{Empty Ochestra Band\\ - Karaoke} &Name&\makecell{Don't You Forget About Me,\\ episode of Supernatural (S11 E12)}&\makecell{Don't You (Forget About Me),\\ original song written and composed\\ by Keith Forsey and Steve Schiff}&Yes\\
    \hline
    bank of nova scotia & \makecell{The Global 2000\\ - Forbes.com} &Company&\makecell{Scotiabank,\\ Canadian bank based in Toronto}&\makecell{Bank of Nova Scotia,\\ bank building in Calgary}&No\\
    \hline
    purple finch & \makecell{The Sea Ranch Association List of Birds} &Common Name&\makecell{Haemorhous purpureus, species of bird}&\makecell{Purple Finch,\\ print in the National Gallery of Art}&No\\
    \hline
    \end{tabular}}}
    \vspace{-10pt}
    \label{tab:EL_case}
\end{table*}

\subsection{Column Type Annotation}
%Column type annotation is the task of identifying the semantic type of a table column, formally defined as below:
We define the task of column type annotation as follow:
% \begin{definition}
% Given a table $T=(C, H, T^C)$ and a set of semantic types $\mathcal{L}$, for column $T^C_{:,j}$, we aim to label it with $l \in \mathcal{L}$ so that all values in $T^C_{:,j}$ have type $l$. \hs{Note that a column can have multiple types.}
% \end{definition}
\begin{definition}
Given a table $T$ and a set of semantic types $\mathcal{L}$, column type annotation refers to the task of annotating a column in $T$ with $l \in \mathcal{L}$ so that all entities in the column have type $l$. Note that a column can have multiple types.
\end{definition}

\nop{Oftentimes $\mathcal{L}$ can be obtained from a KB ontology.} Column type annotation is a crucial task for table understanding and is a fundamental step for many downstream tasks like data integration and knowledge discovery. Earlier work \cite{Mulwad2010UsingLD,Ritze2015MatchingHT,Zhang2017EffectiveAE} on column type annotation often coupled the task with entity linking. First entities in a column  are linked to a KB and then majority voting is employed on the types of the linked entities. More recently, \cite{Chen2018ColNetET,Chen2019LearningSA,Hulsebos2019SherlockAD} have studied column type annotation based on cell texts only. Here we adopt a similar setting, i.e., use the available information in a given table directly for column type annotation without performing entity linking first.
%but also include other elements such as table metadata to better evaluate our model's capability of understanding table content as well as table schema.

\vspace{1mm}
\noindent\textbf{Baselines.} We compare our results with the state-of-the-art model\nop{s, HNN~\cite{Chen2019LearningSA} and} Sherlock~\cite{Hulsebos2019SherlockAD} for column type annotation. \nop{HNN employs a hybrid neural network to extract cell, row and column features.} Sherlock uses 1588 features describing statistical properties, character distributions, word embeddings, and paragraph vectors of the cell values in a column. \nop{both of them build a multi-label classifier based on their respective features?} It was originally designed to predict a single type for a given column. {We change its final layer to $|\mathcal{L}|$ Sigmoid activation functions, each with a binary cross-entropy loss, to fit our multi-label setting.} \revision{}{We also evaluate our model using two datasets in \cite{Chen2019LearningSA}, and include the HNN + P2Vec model as baseline. HNN + P2Vec employs a hybrid neural network to extract cell, row and column features, and combines it with property features retrieved from KB.}

\vspace{1mm}
\noindent\textbf{Fine-tuning \modelnospace.} To predict the type(s) for a column, we first extract the contextualized representation of the column $\mathbf{h}_c$ as follows:
\begin{equation}
        \mathbf{h}_{c} = [\mathtt{MEAN}\left(\mathbf{h}_i^\text{t},\dots\right);\mathtt{MEAN}\left(\mathbf{h}_j^\text{e},\dots\right)].\label{eq:column_repr}
\end{equation}
{Here $\mathbf{h}_i^\text{t}$'s are representations of tokens in the column header, $\mathbf{h}_j^\text{e}$'s are representations of entity cells in the column. The probability of predicting type $l$ is then given as,}
\begin{equation}
    P(l) = \mathrm{Sigmoid}\left(\mathbf{h}_{c}W_{l}+b_{l}\right).
\end{equation}
Same as with the baselines, we optimize the binary cross-entropy loss, $y$ is the ground truth label for type $l$
\begin{equation}
loss =\sum{y\mathrm{log}\left(P(l)\right)+\left(1-y\right)\mathrm{log}\left(1-P(l)\right)}
\end{equation}

\nop{No need to subscript $y$ and $l$ here.}

\vspace{1mm}
\noindent\textbf{Task-specific Datasets.} We refer to Freebase \cite{freebase:datadumps} to obtain semantic types $\mathcal{L}$ {because of its richness, diversity, and scale}. We only keep those columns in our relational table corpus that have at least three linked entities to Freebase, and for each column, we use the common types of its entities as annotations. \nop{After keeping only types that have more than 100 training instances and filtering some redundant types\hs{what are redundant types?}}{We further filter out types with less than 100 training instances and keep only the most representative types. In the end}, we get a total number of 255 types, 628,254 columns from 397,098 tables for training, \nop{13,391  validation columns from 4,844 tables for validation,} 13,025 (13,391) columns from 4,764 (4,844) tables for test (validation). \revision{}{We also test our model on two existing small-scale datasets, T2D-Te and Efthymiou (a subset of WikiGS annotated with types) from \cite{Chen2019LearningSA} and conduct two auxiliary experiments: (1) We first directly test our trained models and see how they generalize to existing datasets. We manually map 24 out of the 37 types used in \cite{Chen2019LearningSA} to our types, which results in 107 (of the original 133) columns in T2D-Te and 416 (of the original 614) columns in Efthymiou. (2) We follow the setting in \cite{Chen2019LearningSA} and use 70\% of T2D as training data, which contains 250 columns.\footnote{\revision{}{We use the data released by \cite{Chen2019LearningSA} (\url{https://github.com/alan-turing-institute/SemAIDA}). The number of instances is slightly different from the original paper. \nop{the author and the paper seem ambiguous. might rephrase and give a citation.}} }}\nop{Note that the tables for training/validation/testing on this task are respectively from the training/validation/testing partitions in our pre-training phase so that there is on overlap between them.}

%We get types $\mathcal{L}$ from Freebase \hs{I think we have more details about how these types are obtained.}.

\vspace{1mm}
\begin{table}[]
    \centering
    \caption{Model evaluation on column type annotation task.}
    \vspace{-10pt}
    \resizebox{0.95\linewidth}{!}{\begin{tabularx}{1.1\linewidth}{|l|Y|Y|Y|}
    \hline
     Method& F1& P&R\\
     \hline
    Sherlock (only entity mention) \cite{Hulsebos2019SherlockAD} &78.47  & 88.40&70.55\\
    {TURL + fine-tuning} (only entity mention)&88.86  &90.54 &87.23\\
    \hline
    {TURL + fine-tuning}  & \textbf{94.75} & \textbf{94.95}&\textbf{94.56}\\
     \hspace{10pt} w/o table metadata  &93.77  &94.80&92.76\\
     \hspace{10pt} w/o learned embedding  &  92.69&92.75&92.63\\
     \hspace{10pt} only table metadata  & 90.24 &89.91&90.58\\
     \hspace{10pt} only learned embedding  & 93.33 &94.72&91.97\\
    \hline
    \end{tabularx}}
    \vspace{-5pt}
    \label{tab:CT}
\end{table}
\begin{table}[]
    \centering
    \caption{\revision{}{Accuracy on T2D-Te and Efthymiou, where {scores for HNN + P2Vec are copied from \cite{Chen2019LearningSA} (trained with 70\% of T2D and Efthymiou respectively\nop{local data (T2D or Efthymiou)} and tested on the rest)}. We directly apply our models by type mapping without retraining. \nop{For all tables, should we add a citation for each of others' models?}} \nop{F\_1 is shown? we never used `ours' now, right? Be consistent with other tables? Also, be a bit more careful with all the new content added.. for example, we used percentage like 94.75 rather than float numbers like 0.966.}}
    \vspace{-10pt}
    \revision{}{\resizebox{0.95\linewidth}{!}{\begin{tabularx}{1.05\linewidth}{|l|Y|Y|}
    \hline
    Method & T2D-Te & Efthymiou \\
    \hline
       HNN + P2Vec (entity mention + KB) \cite{Chen2019LearningSA}  & 0.966 & 0.865\\
       \hline
       TURL + fine-tuning (only entity mention)& 0.888&0.745\\
       \hspace{10pt} + table metadata&0.860 &0.904\\
       \hline
    \end{tabularx}}}
    \vspace{-5pt}
    \label{tab:CT_exisiting}
\end{table}
\begin{table}[]
    \centering
    \caption{\revision{}{Accuracy on T2D-Te and Efthymiou. Here all models use T2D-Tr (70\% of T2D) as training set, {following the setting in \cite{Chen2019LearningSA}}. }}
    \vspace{-10pt}
    \revision{}{\resizebox{0.95\linewidth}{!}{\begin{tabularx}{1.05\linewidth}{|l|Y|Y|}
    \hline
    Method & T2D-Te & Efthymiou \\
    \hline
       HNN + P2Vec (entity mention + KB) \cite{Chen2019LearningSA}  & 0.966 & 0.650\\
       \hline
       TURL + fine-tuning (only entity mention)& 0.940&0.516\\
       \hspace{10pt} + table metadata&0.962 &0.746\\
       \hline
    \end{tabularx}}}
    \vspace{-5pt}
    \label{tab:CT_exisiting_1}
\end{table}
\begin{table*}[]
    \centering
    \caption{Further analysis on column type annotation: {Model performance for 5 selected types}. Results are F1 on validation set.}
    \vspace{-10pt}
    \resizebox{0.95\linewidth}{!}{\begin{tabularx}{\linewidth}{|l|Y|Y|Y|Y|Y|}
    \hline
    Method&person&pro\_athlete&actor&location&citytown\\
    \hline
    Sherlock & 96.85 & 74.39&29.07&91.22&55.72\\
    \hline
    {TURL + fine-tuning}  & \textbf{99.71} & \textbf{91.14}&\textbf{74.85}&\textbf{99.32}&\textbf{79.72}\\
     \hspace{10pt} only entity mention  & 98.44 &87.11&58.86&96.59&60.13\\
     \hspace{10pt} w/o table metadata  & 99.63 &90.38&74.46&99.01&77.37\\
     \hspace{10pt} w/o learned embedding  & 99.38 &90.56&71.39&98.91&75.55\\
     \hspace{10pt} only table metadata  & 98.26 &88.80&70.86&98.11&72.54\\
     \hspace{10pt} only learned embedding  & 98.72 &91.06&73.62&97.78&75.16\\
    \hline
    \end{tabularx}}
    \label{tab:CT_case}
\end{table*}
% \begin{table}[]
%     \centering
%     \caption{Further analysis on column type annotation: {Model performance for 3 selected types}. Results are F1 on validation set.}
%     \vspace{-10pt}
%     \resizebox{0.95\linewidth}{!}{\begin{tabularx}{1.1\linewidth}{|l|Y|Y|Y|}
%     \hline
%     Method&person&pro\_athlete&actor\\
%     \hline
%     Sherlock \cite{Hulsebos2019SherlockAD} & 96.85 & 74.39&29.07\\
%     \hline
%     {TURL + fine-tuning}  & \textbf{99.71} & \textbf{91.14}&\textbf{74.85}\\
%      \hspace{10pt} only entity mention  & 98.44 &87.11&58.86\\
%     %  \hspace{10pt} w/o table metadata  & 99.63 &90.38&74.46\\
%     %  \hspace{10pt} w/o learned embedding  & 99.38 &90.56&71.39\\
%      \hspace{10pt} only table metadata  & 98.26 &88.80&70.86\\
%      \hspace{10pt} only learned embedding  & 98.72 &91.06&73.62\\
%     \hline
%     \end{tabularx}}
%     \vspace{-10pt}
%     \label{tab:CT_case}
% \end{table}

\noindent\textbf{Results.} \revision{}{For the main results on our test set, }{we use the validation set for early stopping in training the Sherlock model, which takes over 100 epochs.}\nop{ while we only fine-tune our model for 10 epochs\hs{if we use validation for early stopping, why still set 10?}.} We evaluate model performance using micro F1, Precision (P) and Recall (R) measures. Results are shown in Table \ref{tab:CT}. 
{Our model substantially outperforms the baseline, even when using the same input information (only entity mention vs Sherlock)}. Adding table metadata information and entity embedding learned during pre-training further boost the performance to $94.75$ under F1. In addition, our model achieves such performance using only 10 epochs for fine-tuning, which demonstrates the efficiency of the pre-training/fine-tuning paradigm. More detailed results for several types are shown in Table \ref{tab:CT_case}, where we observe that all methods work well for coarse-grained types like \texttt{person}\nop{ and location}. However, fine-grained types like \texttt{actor} and \texttt{pro\_athlete} \nop{\texttt{citytown}} are much more difficult to predict. Specifically, it is hard for a model to predict such types for a column only based on entity mentions in cells. On the other hand, using table metadata works much better than using entity mentions (e.g., 70.86 vs 58.86 for \texttt{actor}\nop{ and 72.54 vs 60.13 for citytown}). This indicates the importance of table context information for predicting fine-grained column types.

\revision{}{Results of the auxiliary experiments are summarized in Table \ref{tab:CT_exisiting} and \ref{tab:CT_exisiting_1}. The scores shown are accuracy, i.e., the ratio of correctly labeled columns, given each column is annotated with one ground truth label. For HNN + P2Vec, the scores are directly copied from the original paper \cite{Chen2019LearningSA}. {Note that in Table \ref{tab:CT_exisiting}, the numbers from our models are not directly comparable with HNN + P2Vec, due to mapping the types in the original datasets to ours as mentioned earlier. However, taking HNN + P2Vec trained on in-domain data as reference, we can see that without retraining, our models still obtain high accuracy on both Web table corpus (T2D-Te) and Wikipedia table corpus (Efthymiou)}.
{We also notice that adding table metadata slightly decreases the performance on T2D while increasing that on Efthymiou, which is possibly due to the}\nop{The performance drop on T2D}\nop{comparing what? confusing...88.86 in table 6 with 0.888? I wonder if we need this sentence.. what about the drop on Efthymiou? hints that there is} distributional differences between Wikipedia tables and general web tables. From Table \ref{tab:CT_exisiting_1} we can see that when trained on the same T2D-Tr split, our model with both entity mention and table metadata still outperforms or is on par with the baseline. However, when using only entity mention, \nop{when looking at table 6 and 7/8 together, we don't have learned embedding. Is this clarified somewhere? because we might have them for certain entities, right?}our model does not perform as well as the baseline, especially when generalizing to Efthymiou. This is because: (1) Our model is pretrained with both table metadata and entity embedding. Removing both creates a big mismatch between pretraining and fine-tuning. (2) With only 250 training instances, it is easy for deep models to overfit. The better performance of models leveraging table metadata under both settings demonstrates the usefulness of context for table understanding.}

\subsection{Relation Extraction}
\begin{table}[]
    \centering
    \caption{Model evaluation on relation extraction task.}
    \vspace{-10pt}
    \resizebox{\linewidth}{!}{\begin{tabularx}{1.1\linewidth}{|l|Y|Y|Y|}
    \hline
     Method& F1& P&R\\
    \hline
    BERT-based & 90.94 & 91.18 &90.69\\
    {TURL + fine-tuning} (only table metadata)& 92.13  &91.17&93.12\\
    \hline
    {TURL + fine-tuning}  & \textbf{94.91} & \textbf{94.57} &\textbf{95.25}\\
     \hspace{10pt} w/o table metadata  & 93.85 &93.78&93.91\\
     \hspace{10pt} w/o learned embedding  & 93.35 &92.90&93.80\\
    \hline
    \end{tabularx}}
    \vspace{-5pt}
    \label{tab:RE}
\end{table}

\begin{table}[]
    \centering
    \caption{Relation extraction results of an entity linking based system, under different agreement ratio thresholds. }
    \vspace{-10pt}
    \revision{}{\resizebox{0.6\linewidth}{!}{\begin{tabular}{|c|c|c|c|}
    \hline
        Min Ag. Ratio & F1&P&R \\
    \hline
        0 & 68.73 &60.33&\textbf{79.85}\\
        0.4 & 82.10&94.65&72.50\\
        0.5 & 77.68&98.33&64.20\\
        0.7 & 63.10&99.37&46.23\\
    \hline
    \end{tabular}}}
    \vspace{-10pt}
    \label{tab:RE_entity}
\end{table}
% \begin{table*}[]
%     \centering
%     \begin{tabularx}{0.95\linewidth}{|l|Y|Y|Y|Y|Y|}
%     \hline
%     Method&directed\_by&written\_by&nationality&award\_winner&award\_nominee\\
%     \hline
%     BERT&98.84&60.47&97.40&63.64&68.02\\
%     \hline
%     \add{TURL + fine-tuning}  & \textbf{99.14} & \textbf{74.16}&98.42&\textbf{64.58}&  \textbf{79.32}\\
%      \hspace{10pt}- w/o table metadata  & 98.84 &71.43&98.42&61.70&75.47\\
%      \hspace{10pt}- w/o learned embedding  & 98.84 &67.37&\textbf{99.68}&58.82&72.20\\
%      \hspace{10pt}- only table metadata  & 98.56 &70.97&98.71&62.96&69.05\\
%     \hline
%     \end{tabularx}
%     \caption{Further analysis on relation extraction: Model performance for 5 selected relations. Results are F1 on validation set.}
%     \label{tab:RE_case}
% \end{table*}

Relation extraction is the task of mapping column pairs in a table to relations in a KB. A formal definition is given as follows.

\begin{definition}
Given a table $T$ and a set of relations $\mathcal{R}$ in KB. For a subject-object column pair in $T$, we aim to annotate it with $r \in \mathcal{R}$ so that $r$ holds between all entity pairs in the columns. 
\end{definition}

Most existing work \cite{Mulwad2010UsingLD,Ritze2015MatchingHT,Zhang2017EffectiveAE} assumes that all relations between entities are known in KB and relations between columns can be easily inferred based on entity linking results. However, such methods rely on entity linking performance and suffer from KB incompleteness. Here we aim to conduct relation extraction without explicitly linking table cells to entities. \textit{This is important as it allows the extraction of new knowledge from Web tables for tasks like knowledge base population.}

\vspace{1mm}
\noindent\textbf{Baselines.} We compare our model with a state-of-the-art text based relation extraction model~\cite{zhang2019ernie} which utilizes a pretrained BERT model to encode the table information. {For text based relation extraction, the task is to predict the relation between two entity mentions in a sentence. Here we adapt the setting by treating the concatenated table metadata as a sentence, and the headers of the two columns as entity mentions.}\nop{so we don't use the cell texts?} \nop{cite source model here?}
\revision{}{Although our setting is different from the entity linking based relation extraction systems in \cite{Mulwad2010UsingLD,Ritze2015MatchingHT,Zhang2017EffectiveAE}, here we implement a similar system using our entity linking model described in Section \ref{sec:EL}, and obtain relation annotations based on majority voting of linked entity pairs, i.e., predict a relation if it holds between a minimum portion of linked entity pairs in KB (i.e., the minimum agreement ratio is larger than a threshold). }

\vspace{1mm}
\noindent\textbf{Fine-tuning \modelnospace.} We use similar model architecture as column type annotation as follows.
\begin{equation}
    P(r) = \mathrm{Sigmoid}\left([\mathbf{h}_{c};\mathbf{h}_{c'}]W_r+b_r\right).
\end{equation}
\nop{$c1, c2 \rightarrow c, c'$?}

Here $\mathbf{h}_{c}, \mathbf{h}_{c'}$ are aggregated representation for the two columns obtained same as Eqn. \ref{eq:column_repr}. We use binary cross-entropy loss for optimization. \nop{check notations..}

\begin{figure}
    \centering
    \includegraphics[width=0.7\linewidth]{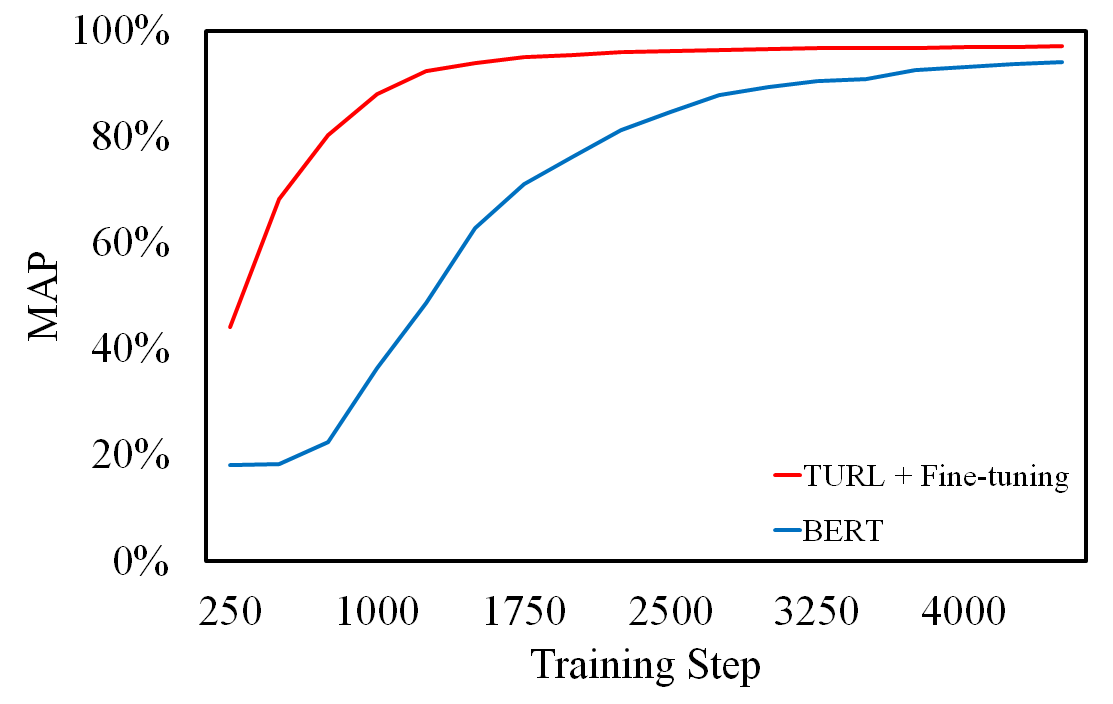}
    \vspace{-10pt}
    \caption{Comparison of fine-tuning our model and BERT for relation extraction: Our model converges much faster.\nop{ Here one training step refers to training for one batch.} \nop{what does the "training steps" mean? we only mentioned epochs in text.}}
    \vspace{-15pt}
    \label{fig:RE_CMP}
\end{figure}

\noindent\textbf{Task-specific Datasets.} {We prepare datasets for relation extraction in a similar way as the previous column type annotation task, based on our pre-training table partitions. Specifically, we obtain relations $\mathcal{R}$ from Freebase. For each table in our corpus, we pair its subject column with each of its object columns, and annotate the column pair with relations shared by more than half of the entity pairs in the columns. We only keep relations that have more than 100 training instances.} Finally, we obtain a total number of 121 relations, 62,954 column pairs from 52,943 tables for training, and 2072 (2,175) column pairs from 1467 (1,560) tables for test (validation). %\hs{so, on this task, we don't need to run on the datasets pointed out by the reviewers?}

\vspace{1mm}
\noindent\textbf{Results.} We fine-tune the BERT-based model for 25 epochs. We use micro F1, Precision (P) and Recall (R) measures for evaluation. Results are summarized in Table \ref{tab:RE}.

From Table \ref{tab:RE} we can see that: (1) Both the BERT-based baseline and our model achieve good performance, with F1 scores larger than 0.9. (2) Our model outperforms the BERT-based baseline under all settings, even when using the same information (i.e., only table metadata vs BERT-based). Moreover, we plot the mean average precision (MAP) curve on our validation set during training in Figure \ref{fig:RE_CMP}. As one can see, our model converges much faster in comparison to the BERT-based baseline, demonstrating that our model learns a better initialization through pre-training. \nop{Some detailed results for several relations are shown in Table \ref{tab:RE_case}.\hs{do we have some analysis here? may remove this table if no insights.}}

\revision{}{As mentioned earlier, we also experiment with an entity linking based system. Results are summarized in Table \ref{tab:RE_entity}. We can see that it can achieve high precision, but suffers from low recall: The upper-bound of recall is only 79.85\%, achieved at an agreement ratio of 0 (i.e., taking all relations that exist between the linked entity pairs as positive). {As seen from Table \ref{tab:RE} and \ref{tab:RE_entity}, our model also substantially outperforms the system based on a strong entity linker.} }

\subsection{Row Population}
Row population is the task of augmenting a given table with more rows or row elements. For relational tables,\nop{ since each row is centered around a subject entity (or core entity), with other columns describing attributes of that entity,} existing work has tackled this problem by retrieving entities to fill the subject column \cite{Zhang2020WebTE,zhang2017entitables}. \nop{This is similar to the problem of concept expansion, where we want to extend a set of seed entities with additional entities belong to the same concept \cite{wang2015concept,wang2008iterative,bron2013example,he2011seisa}. The key difference here is that in row population, given none or few seed entities, it is important for the model to understand information given by table metadata.} A formal definition of the task is given below.

\begin{definition}
Given a partial table $T$, and an optional set of seed subject entities, row population aims to retrieve more entities to fill the subject column.
\end{definition}

\noindent\textbf{Baselines.} We adopt models from \cite{zhang2017entitables} and \cite{Deng2019Table2VecNW} as baselines. \cite{zhang2017entitables} uses a generative probabilistic model which ranks candidate entities\nop{based on the multi-conditional probability} considering both table metadata and entity co-occurrence statistics. \cite{Deng2019Table2VecNW} further improves upon \cite{zhang2017entitables} by utilizing entity embeddings trained on the table corpus to estimate entity similarity. We use the same candidate generation module from \cite{zhang2017entitables} for all methods, which formulates a search query using either the table caption or seed entities and then retrieves tables via the BM25 retrieval algorithm. Subject entities in those retrieved tables will be candidates for row population.

\vspace{1mm}
\noindent\textbf{Fine-tuning \modelnospace.} {We adopt the same candidate generation module used by baselines}\nop{what does this mean? to generate ``candidates''}. {We then append the [MASK] token to the input, and use the hidden representation $\mathbf{h}^\text{e}$ of [MASK] to rank these candidates as shown in Table \ref{tab:task_overview}. We fine-tune our model with multi-label soft margin loss as shown below:
\begin{equation}
\begin{aligned}
    P(e) &= \mathrm{Sigmoid}\left(\mathtt{LINEAR}(\mathbf{h}^\text{e})\cdot\mathbf{e}^\text{e}\right),\\
    loss &= \sum_{e\in \mathcal{E}_C}{y\mathrm{log}\left(P(e)\right)+\left(1-y\right)\mathrm{log}\left(1-P(e)\right).}
\end{aligned}
\end{equation}
Here $\mathcal{E}_C$ is the candidate entity set, and $y$ is the ground truth label of whether $e$ is a subject entity of the table.}
\nop{, similarly to Masked Entity Recovery in the pre-training phase. \hs{do we have new parameters to learn? using ranking loss? here you have multiple entities to rank? there is only one [mask] which is unlike MER that has entity mention and entity embedding?}}

\vspace{1mm}
\noindent\textbf{Task-specific Datasets.} Tables in our pre-training set with more than 3 subject entities are used for {fine-tuning \model and developing baseline models}, while tables in our held-out set with more than 5 subject entities are used for evaluation. In total, we obtain 432,660 tables for fine-tuning with 10 subject entities on average, and 4,132 (4,205) tables for test (validation) with 16 (15) subject entities on average.

% We use a generative probabilistic model from \cite{zhang2017entitables} as our baseline. It ranks candidate entities based on the multi-conditional probability $P(e|E,L,c)$ which is calculated as follows:
% \begin{align}
%     P(e|E,L,c) &= \frac{P(E|e)P(L|e)P(c|e)P(e)}{P(E)P(L)P(c)}\nonumber \\
%               &\propto P(e|E)P(L|e)P(c|e)
% \end{align}
% Here $P(e|E)$ is the posterior probability that expresses entity similarity. For baseline model, we estimate entity similarity using the co-occurrence statistics from the table corpus:
% \begin{equation}
%     P_{TC}(e|E)=\frac{\#(e,E)}{\#(E)},
% \end{equation}
% where $\#(e,E)$ is the number of tables that contains the candidate entity together with all seed entities, and $\#(E)$ is the number of tables that contain all seed entities.

% $P(L|e)$ is the column labels likelihood and P(c|e) is the caption likelihood.
\vspace{1mm}
\noindent\textbf{Results.}
\nop{We fine-tune our model for 10 epochs. \hs{could we move this sentence somewhere else if we always use 10 epochs?}}{ The experiments are conducted under two settings: without any seed entity and with one seed entity. For experiments without the seed entity, we only use table caption for candidate generation. For entity ranking in EntiTables \cite{zhang2017entitables}, we use the combination of caption and label likelihood when there is no seed entity, and only use entity similarity when seed entities are available. This strategy works best\nop{they work best} on our validation set. As shown in Table \ref{tab:RP}, our method outperforms all baselines. In particular, previous methods rely on entity similarity and are not applicable or have poor results when there is no seed entity available. Our method achieves a decent performance even without any seed entity, which demonstrates the effectiveness of \model for generating contextualized representations based on both table metadata and content.} \nop{why ours don't have recall in table 10?}
\begin{table}[]
    \centering
    \caption{Model evaluation on row population task. {Recall is the same for all methods because they share the same candidate generation module.}}
    \vspace{-10pt}
    \resizebox{0.75\linewidth}{!}{\begin{tabular}{|l|c|c|c|c|}
    \hline
    \# seed &\multicolumn{2}{c|}{0}&\multicolumn{2}{c|}{1}\\\hline
    Method & MAP & Recall& MAP & Recall\\\hline
    EntiTables~\cite{zhang2017entitables} & 17.90 &63.30&42.31&78.13\\
    Table2Vec~\cite{Deng2019Table2VecNW} &-  &63.30&20.86&78.13\\
    {TURL + fine-tuning}  & \textbf{40.92} &63.30&\textbf{48.31}&78.13\\
    \hline
    \end{tabular}}
    \vspace{-5pt}
    \label{tab:RP}
\end{table}
\begin{table}[]
    \centering
    \caption{Model evaluation on cell filling task.}
    \vspace{-10pt}
   \begin{tabularx}{0.75\linewidth}{|l|Y|Y|Y|Y|}
    \hline
    Method& P @ 1 & P @ 3 & P @ 5 & P @ 10 \\ \hline
    Exact & 51.36 & 70.10& 76.80 & 84.93\\ 
    H2H & 51.90&  70.95& 77.33& 85.44 \\ 
    H2V & 52.23& 70.82 &77.35 & 85.58 \\ \hline
    {TURL} &\textbf{54.80} & \textbf{76.58} &\textbf{83.66} & \textbf{90.98} \\ \hline
    \end{tabularx}
    \vspace{-5pt}
    \label{tab:CF}
\end{table}
\begin{table}[]
    \centering
    \caption{Model evaluation on schema augmentation task.}
    \vspace{-10pt}
    \resizebox{0.75\linewidth}{!}{\begin{tabularx}{0.8\linewidth}{|l|Y|Y|}
\hline
\multirow{2}{*}{\textbf{Method}} & \multicolumn{2}{c|}{\#seed column labels} \\ \cline{2-3} 
 & 0 & 1\\ \hline
kNN & 80.16 & \textbf{82.01}\\ 
{TURL + fine-tuning}&\textbf{81.94} & 77.55 \\\hline
\end{tabularx}}
    \vspace{-10pt}
    \label{tab:SA}
\end{table}

\begin{table*}[]
    \centering
    \caption{Case study on schema augmentation. {Here we show average precision (AP) for each example. Support Caption is the caption of the source table that kNN found to be most similar to the query table.} {Our model performs worse when there exist source tables that are very similar to the query table (e.g., comparing support caption vs query caption).} \nop{I feel `AP' and `support caption' might be unclear.} }
    \vspace{-10pt}
    \resizebox{0.95\linewidth}{!}{\begin{tabular}{|c|c|c|c|c|c|c|}
    \hline
  Method&{Query} Caption&Seed&Target&AP&Predicted&Support Caption\\
    \hline
    kNN&\multirow{2}{*}{2010 santos fc season out}&\multirow{2}{*}{pos.}&\multirow{2}{*}{\makecell{name,\\ moving to}}&1.0&\makecell{moving to, name, player,\\ moving from, to}&2007 santos fc season out\\ \cline{5-7}
    Ours&&&&0.58&\makecell{moving to, fee/notes,\\ destination club, fee, loaned to}&-\\\hline
    kNN&\multirow{2}{*}{\makecell{first ladies and gentlemen\\ of panama list}}&\multirow{2}{*}{no.}&\multirow{2}{*}{\makecell{name,\\ president}}&0.20&\makecell{country, runner-up,\\ champion, player, team team}&\makecell{first ladies of chile\\ list of first ladies}\\ \cline{5-7}
    Ours&&&&0.14&\makecell{year, runner-up, spouse,\\ name, father}&-\\\hline
    kNN&\multirow{2}{*}{\makecell{list of radio stations in\\ metro manila am stations}}&\multirow{2}{*}{name}&\multirow{2}{*}{\makecell{format,\\ covered location}}&1.0&\makecell{format, covered location, company,\\ call sign, owner}&\makecell{list of radio stations in\\ metro manila fm stations}\\ \cline{5-7}
    Ours&&&&0.83&\makecell{format, owner, covered location,\\ city of license, call sign}&-\\\hline
    \end{tabular}}
    \vspace{-10pt}
    \label{tab:SA_error}
\end{table*}
\subsection{Cell Filling}
We examine the utility of our model in filling other table cells, assuming the subject column is given. \nop{This is similar to the entity augmentation task in InfoGather \cite{yakout2012infogather}, which assumes subject entities and aims to find the corresponding object entities given a column name (\textit{augmentation by attribute name}) or  given a subject-object entity pair (\textit{augmentation by example}). Another variant of the cell filling task is to predict the missing values in a table, also known as data imputation \cite{zhang2019auto,ahmadov2015towards}. Data imputation focuses more on individual cells and considers cells containing numerical value, date other than entity.} This is similar to the setting in \cite{zhang2019auto, yakout2012infogather}, which we formally define as follows. 

\begin{definition}
Given a partial table $T$ with the subject column filled and an object column header, cell filling aims to predict the object entity for each subject entity.
\end{definition}

\noindent\textbf{Baselines.} We adopt \cite{zhang2019auto} as our base model. It has two main components, candidate value finding and value ranking. The same candidate value finding module is used for all methods: Given a subject entity $e$ and object header $h$ for the to-be-filled cells, we find all entities that appear in the same row with $e$ in our {pre-training} table corpus \nop{needs to point out it's pre-training table corpus?}, and only keep entities whose {source} header $h'$ is related to $h$. Here we use the formula from \cite{zhang2019auto} to measure the relevance of two headers $P(h'|h)$,
\begin{equation}
    P(h'|h) = \frac{n(h',h)}{\sum_{h''}n(h'',h)}.
\end{equation}
Here $n(h',h)$ is the number of table pairs in the table corpus that contain the same entity for a given subject entity in columns $h'$ and $h$. The intuition is that if two tables contain the same object entity for a given subject entity $e$ in columns with headings $h_a$ and $h_b$, then $h_a$ and $h_b$ might refer to the same attribute. For value ranking, the key is to match the given header $h$ with the source header $h'${, we can then get the probability of the candidate entity $e$ belongs to the cell $P(e|h)$ as follows:
\begin{equation}
    P(e|h)=\mathrm{MAX}\left(\mathrm{sim}(h',h)\right).
\end{equation}
Here $h'$'s are the source headers associated with the candidate entity in the pre-training table corpus. $\mathrm{sim}(h',h)$ is the similarity between $h'$ and $h$.} We develop three baseline methods for $\mathrm{sim}(h',h)$: (1) \textbf{Exact}: predict the entity with exact matched header, (2) \textbf{H2H}: use the $P(h'|h)$ described above. (3) \textbf{H2V}: similar to \cite{Deng2019Table2VecNW}, we train header embeddings with Word2Vec on the table corpus. We then measure the similarity between headers using cosine similarity.\nop{to check this part. A bit unclear.}

\vspace{1mm}
\noindent\textbf{Fine-tuning \modelnospace.} {Since cell filling is very similar to the MER pre-training task, we do not fine-tune the model, and directly use [MASK] to select from candidate entities same as MER (Eqn. \ref{eq:e_prob}).}

\vspace{1mm}
\noindent\textbf{Task-specific Datasets.}
 To evaluate different methods on this task, we use the held-out test tables in our pre-training phase and extract from them those subject-object column pairs that have at least three valid entity pairs. Finally we obtain 9,075 column pairs for evaluation.

\vspace{1mm}
\noindent\textbf{Results.}
For candidate value finding, using all entities appearing in the same row with a given subject entity $e$ achieves a recall of 62.51\% with 165 candidates on average. After filtering with $P(h'|h)>0$, the recall drops {slightly} to 61.45\% and the average number of candidates reduces to 86. For value ranking, we only consider those test instances with the target object entity in the candidate set and evaluate them under Precision@K (or, P@K). Results are summarized in Table \ref{tab:CF}, from which we show: (1) Simple \textbf{Exact} match achieves decent performance, and using \textbf{H2H} or \textbf{H2V} only sightly improves the results. (2) Even though our model directly ranks the candidate entities without explicitly using their source table information, it outperforms other methods. This indicates that our model already encodes the factual knowledge in tables into entity embeddings through pre-training. 

\subsection{Schema Augmentation}
Aside from completing the table content, another direction of table augmentation focuses on augmenting the table schema, i.e., discovering new column headers to extend a table with more columns \cite{cafarella2008webtables, yakout2012infogather, zhang2017entitables, Deng2019Table2VecNW}.\nop{ Existing methods differ in terms of what table components are used as input. Some work \cite{cafarella2008webtables, lehmberg2015mannheim} uses seed headers only, while most others \cite{yakout2012infogather,zhang2017entitables,Deng2019Table2VecNW} uses all table components. In this paper, we follow the latter and define the task as below.} {Following \cite{yakout2012infogather,zhang2017entitables,Deng2019Table2VecNW}, we formally define the task below.}
\begin{definition}
Given a partial table $T$, which has a caption and zero or a few seed headers, {and a header vocabulary $\mathcal{H}$}, schema augmentation aims to recommend a ranked list of headers $h \in \mathcal{H}$ to add to $T$.
\end{definition}

\nop{
\ww{Question: for the 1-seed-header setting, is it always the subject column as seed?  Predicting 1st column vs. 2nd/3rd/4th ... seems different---one is understanding the topic of the table, while the other is predicting the salient attributes of the subject within the context.  Also worth inspecting how many subject columns have dummy headers such as "Name"---these are not worth including in evaluation.}
\xd{Right now I always use the first column header, which is the subject column for the most of the time. I agree with the difference between using seed header and without seed header. The current schema augmentation result is still not good, so I put some error analysis here, will try to better justify it}
}

\noindent\textbf{Baselines.} We adopt the method in \cite{zhang2017entitables} which searches {our pre-training} table corpus for related tables, and use headers in those related tables for augmentation. More specifically, we encode the given table caption as a tf-idf vector\nop{do you consider seed headers here?} and then use the K-nearest neighbors algorithm (kNN) \cite{Altman1992AnIT} with cosine similarity to find the top-10 most related tables. {We rank headers from those tables by aggregating the cosine similarities for tables they belong to. When seed headers are available, we re-weight the tables by the overlap of their schemas with seed headers same as \cite{zhang2017entitables}.} \nop{The phrasing here needs to be more precise.  The entire process of finding $k$ nearest neighbors and weighted aggregation can be considered as an instance of $k$NN regression, but the action of finding $k$ nearest neighbors alone does not involve $k$NN---instead, it's preparing the input for any $k$NN variant.}

\vspace{1mm}
\noindent\textbf{Fine-tuning \modelnospace.} \nop{We adjust the MLM pre-training task for schema augmentation. We mask out the headers and use the output for [MASK] to recover them in a given header vocabulary $\mathcal{H}$. \hs{what if there are multiple tokens in the header?}}{We concatenate the table caption, seed headers and a [MASK] token as input to our model. The output for [MASK] is then used to predict the headers in a given header vocabulary $\mathcal{H}$. We fine-tune our model use binary cross-entropy loss.}

\vspace{1mm}
\noindent\textbf{Task-specific Datasets.} We collect $\mathcal{H}$ from the pre-training table corpus. We normalize the headers using simple rules, only keep those that appear in at least 10 different tables, and finally obtain 5652 unique headers, with 316,858 training tables and 4,646 (4,708) test (validation) tables.

\vspace{1mm}
\noindent\textbf{Results.}
{We fine-tune our model for 50 epochs for this task, based on the performance on the validation set.} We use mean average precision (MAP) for evaluation. 

From Table \ref{tab:SA}, we observe that both kNN baseline and our model achieve good performance\nop{what is the metric? overall need to clarify how measures are computed in all tasks.}. Our model works better when no seed header is available, but does not perform as well when there is one seed header. We then conduct a further analysis in Table \ref{tab:SA_error} using a few examples: One major reason why kNN works well is that there exist tables in the pre-training table corpus that are very similar to the query table and have almost the same table schema. On the other hand, our model oftentimes suggests plausible, semantically related headers, but misses the ground-truth headers. %sometimes predicts synonyms of the target header or

\begin{figure}
    \centering
    \begin{subfigure}{\linewidth}
    \centering
    \includegraphics[width=0.8\linewidth]{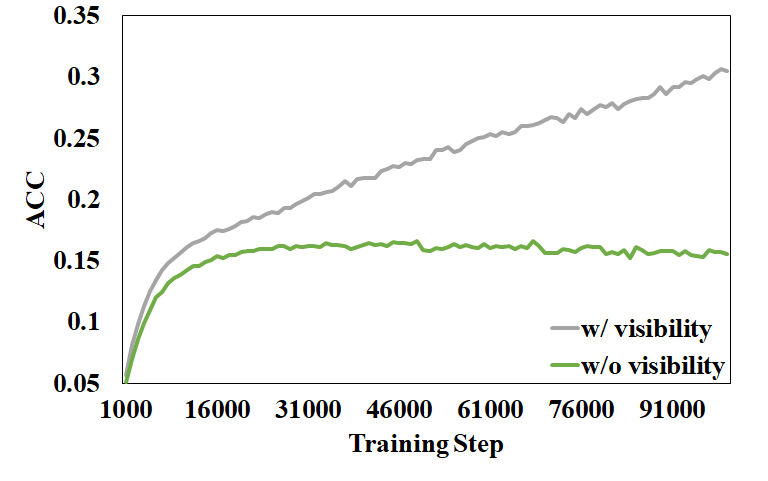}
    \vspace{-5pt}
    \caption{Effect of visibility matrix.}
    \label{fig:vis_acc}
    \end{subfigure}
    \begin{subfigure}{\linewidth}
    \centering
    \includegraphics[width=0.8\linewidth]{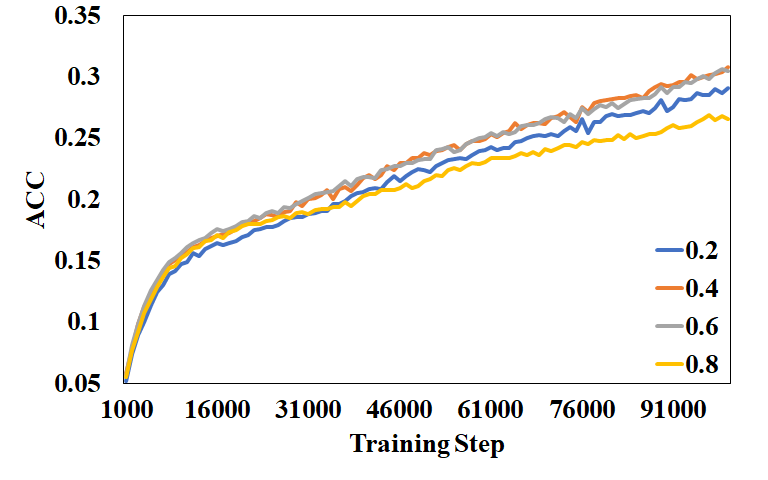}
    \vspace{-5pt}
    \caption{Effect of different MER mask ratios.}
    \label{fig:ratio_acc}
    \end{subfigure}
    \vspace{-5pt}
    \caption{Ablation study results.}
    \vspace{-12pt}
    \label{fig:ablation}
\end{figure}
% \begin{figure}
%     \centering
%     \includegraphics[width=0.8\linewidth]{figures/ratio_acc.png}
%     \caption{Effect of different MER mask ratios.}
%     \label{fig:ratio_acc}
% \end{figure}

\subsection{Ablation Study}
In this section, we examine the effects of two important designs in \modelnospace: the visibility matrix and MER with different mask ratios. {During the pre-training phase, at each training step, we evaluate \model on the validation set for \textit{object entity prediction}. We choose this task because it is similar to the cell filling downstream task and it is convenient to conduct during pre-training (e.g., ground-truth is readily available and no need to modify the model architecture, etc.)}. % subject entity retrieval to justify why we don't use other downstream tasks to validate. 

{Given a table in our validation set, \nop{(1) for subject entity retrieval, we concatenate the table caption and subject column header, and append a [MASK] entity cell as input to \model, which will produce a contextualized representation for [MASK] and use it to rank candidate entities via Eqn. \ref{eq:e_prob} to fill the subject column. We calculate MAP to evaluate the ranked entities w.r.t. the ground-truth entities in the table; (2)} we predict each object entity by first masking the entity cell {(both $\mathbf{e}^\text{e}$ and $\mathbf{e}^\text{m}$)} and obtaining a contextualized representation of the [MASK] (which attends to the table caption, corresponding header, as well as other entities in the same row/column before the current cell position\nop{double check..}) and then applying Eqn. \ref{eq:e_prob}.} We compare the top-1 predicted entity with the ground truth and show the accuracy (ACC) on average. Results are summarized in Figure \ref{fig:ablation}. 

%\vspace{1mm}
{Figure \ref{fig:vis_acc} clearly demonstrates the advantage of our visibility matrix design. Without the visibility matrix (an element can attend to every other element during pre-training), it is hard for the model to capture the most relevant information (e.g., relations between entities) in the table for prediction. From Figure \ref{fig:ratio_acc}, we observe that at a mask ratio of 0.8, the objective entity prediction performance drops in comparison with other lower ratios. This is because this task requires the model to not only understand the table metadata, but also learn the relation between entities. A high mask ratio forces the model to put more emphasis on the table metadata, while a lower mask ratio encourages the model to leverage the relation between entities. {Meanwhile, a very low mask ratio such as 0.2 also hurts the pre-training performance, because only a small portion of entity cells are actually used for training in each iteration. A low mask ratio also creates a mismatch between pre-training and fine-tuning, since for many downstream tasks, only few seed entities are given.} Considering both aspects as well as that the results are not sensitive w.r.t. this parameter, we set the MER mask ratio at 0.6 in pre-training.}

\nop{For comparison, we simulate the row population and cell filling task on the validation set \hs{why do we choose these two? or choose to do so?}. More specifically, we append [MASK] token to the input entity sequence \hs{does the input sequence mean all entities or only entities in a subject column?} and pair each entity with a [MASK] token \hs{do you pair with the previously added [MASK] token?}. The former [MASK] can only see the table caption and subject column header and is used to rank entities for row population. While the latter [MASK] is used to recover the paired entity and simulate cell filling. \hs{do you mean for the latter [MASK]? if so, does not make sense to me: It has the same visibility as the paired entity except that it cannot see the paired entity and all entities after it.}}

\nop{\hs{From Figure \ref{fig:cmp_ratio}, our interesting observations include: (1) At a mask ratio of 0.8, the subject entity retrieval performance improves while the objective entity prediction performance drops in comparison with other lower ratios. The phenomenon makes sense because: For the subject entity retrieval task, since a masked entity only sees the table metadata, the model will focus more on understanding the topic of the table to retrieve the correct set of subject entities. On the other side, cell filling requires the model to not only understand the table metadata, but also know the relation between entities. A high mask ratio forces the model to put more emphasis on the table metadata, while a lower mask ratio encourages the model to better understand the relation between entities. \add{(2) worth talking about the ration of 0.2, as it is interesting as well - both do not work well as I guess the model does not learn much about relational knowledge when there is not much to predict} Considering both settings, we set the MER mask ratio at 0.6 in pre-training.}}
%as we want to the model encode as much relational knowledge about entities as possible. and leave the further improvement of row population performance to fine-tuning phase
%From Figure \ref{fig:cmp_ratio}, we can see that there is a trade-off between row population and cell filling performance during pre-training. With a higher mask ratio, the row population performance improves while the cell filling performance drops. The reason is that for the simulated row population task, since it only sees the table metadata, it focuses more on validating the model's capability to understand the topic of the table and retrieve the correct set of subject entities. On the other side, cell filling requires the model to not only interpret the table metadata, but also know the relation between entities. A higher mask ratio forces the model to put more emphasis on the table metadata, while a lower mask ratio encourages the model to better understand the relation between entities. For pre-training, we choose the mask ratio for MER as 0.6, as we want to the model encode as much relational knowledge about entities as possible. And we leave the further improvement of row population performance to fine-tuning phase.

\nop{Figure \ref{fig:cmp_vis} demonstrates the advantage of our visibility matrix design. Without table structure information, it is hard for the model to extract factual knowledge about entities in the table. We can see that pre-training without visibility matrix has poor performance on the simulated cell filling task, as the model fails to capture the relations between entities. We also notice that the model without visibility matrix still works well on row population, which proves our assumption above that row population only focuses on understanding the table topic while neglects the entity relations. \hs{but without visibility, a subject entity cell will also see others in the same column?}}
\section{Conclusion}
{This paper presents a novel pre-training/fine-tuning framework (\modelnospace) for relational table understanding. It consists of a structure-aware Transformer encoder to model the row-column structure as well as a new Masked Entity Recovery objective to capture the semantics and knowledge in relational Web tables during pre-training. On our compiled benchmark, we show that \model can be applied to a wide range of tasks with minimal fine-tuning and achieves superior performance {in most scenarios}. {Interesting future work includes: (1) Focusing on other types of knowledge such as numerical attributes in relational Web tables, in addition to entity relations\nop{outside of factual knowledge about entities,}. (2) Incorporating the rich information contained in an external KB into pre-training.} %is also an interesting direction
}
\nop{we compile a benchmark that consists of 6 different tasks for table understanding.
To systematically evaluate \model, we compile a benchmark that consists of 6 different tasks for table understanding (e.g., relation extraction, cell filling). We show that \model generalizes well to all these tasks and substantially outperforms existing methods \add{in most cases}. Our source code, benchmark, as well as pre-trained models are available online to facilitate future research.}

\begin{acks}
We would like to thank the anonymous reviewers for their helpful comments. Authors at the Ohio State University were sponsored in part by Google Faculty Award, the Army Research Office under cooperative agreements W911NF-17-1-0412, NSF Grant IIS1815674, NSF CAREER \#1942980, Fujitsu gift grant, and Ohio Supercomputer Center \cite{OhioSupercomputerCenter1987}. The views and conclusions contained herein are those of the authors and should not be interpreted as representing the official policies, either expressed or implied, of the Army Research Office or the U.S. Government. The U.S. Government is authorized to reproduce and distribute reprints for Government purposes notwithstanding any copyright notice herein.
\end{acks}
\bibliographystyle{ACM-Reference-Format}
\balance
\bibliography{bibliography.bib}

% \newpage
% \nobalance
% \appendix
% \input{sections/response}

\end{document}